\documentclass{article}
\usepackage{format}
\usepackage{latexsym}
\usepackage[english]{babel}
\usepackage{bm}
\usepackage[T1]{fontenc}
\usepackage[utf8]{inputenc}
\usepackage{ae}
\usepackage{graphicx}
\usepackage{subcaption}
\usepackage[graphicx]{realboxes}
\usepackage{epsfig}
\usepackage{color}
\usepackage{pgf}
\usepackage{pstricks,pst-grad}
\usepackage{float}
\usepackage{sidecap}
\sidecaptionvpos{figure}{c}
\usepackage[normalem]{ulem}
\usepackage{latexsym}
\usepackage{amsfonts}
\usepackage{amssymb}
\usepackage{amsthm}
\usepackage{amscd}
\usepackage{amsmath}
\usepackage{mathrsfs}
\usepackage{breqn}
\usepackage[bbgreekl]{mathbbol}
\usepackage{mathtools}
\usepackage{color}
\usepackage{colortbl}
\usepackage{framed}
\usepackage{tcolorbox}
\usepackage{verbatim}
\usepackage{pdfpages}

\usepackage{microtype}
\usepackage[inline]{enumitem}
\usepackage{verbatim}
\usepackage{placeins}
\usepackage[full]{textcomp}
\usepackage{wasysym}
\usepackage{appendix}
\usepackage[nolist,nohyperlinks]{acronym}
\usepackage{diagbox}
\usepackage{booktabs}
\usepackage{xr}
\usepackage{amsthm}
\usepackage{qcircuit}
\usepackage{physics}

\usepackage{hyperref}
\hypersetup{
    colorlinks=true,
    citecolor=black,
    linkcolor=red,
    linkbordercolor=blue,
}

\newcommand{\g}{\text{glob}}

\newcommand{\inst}{\text{inst}}

\begin{acronym}[MPTPS]
    \acro{QVAE}{quantum variational autoencoder}
    \acro{QML}{quantum machine learning}
    \acro{QFL}{quantum federated learning}
    \acro{QNN}{quantum neural network}
    \acro{VAE}{variational autoencoder}
    \acro{PSD}{positive semi-definite}
    \acro{PCC}{Pearson-Correlation coefficient}
    \acro{SNPs}{single nucleotide polymorphisms}
    \acro{pdf}{probability density function}
\end{acronym}
\usepackage{csquotes}
\usepackage[backend=biber,sorting=none]{biblatex}
\begin{document}

\renewcommand{\figureautorefname}{Fig.}
\renewcommand{\equationautorefname}{Eq.}
\renewcommand{\sectionautorefname}{Sec.}
\renewcommand{\subsectionautorefname}{subsec.}
\renewcommand{\tableautorefname}{Tab.}

\title{Efficient Privacy-Preserving Training of Quantum Neural Networks by Using Mixed States to Represent Input Data Ensembles}

\author{%
Gaoyuan Wang$^{1,2}$ \quad Jonathan Warrell$^{1,2}$ \quad Mark Gerstein$^{1,2,3,4,5*}$ \\ \\
$^1$ Program in Computational Biology and Bioinformatics,\quad \\
$^2$ Department of Molecular Biophysics and Biochemistry,\quad  \\
$^3$ Department of Computer Science,\quad  \\
$^4$ Department of Statistics \& Data Science,\quad  \\
$^5$ Department of Biomedical Informatics \& Data Science,\quad  \\
Yale University, New Haven, Connecticut 06520, USA \quad  \\
\texttt{*Corresponding author: pi@gersteinlab.org}\\
}
\maketitle

\begin{abstract}

Quantum neural networks (QNNs) are gaining increasing interest due to their potential to detect complex patterns in data by leveraging uniquely quantum phenomena. This makes them particularly promising for biomedical applications. In these applications and in other contexts, increasing statistical power often requires aggregating data from multiple participants. However, sharing data, especially sensitive information like personal genomic sequences, raises significant privacy concerns.
Quantum federated learning offers a way to collaboratively train QNN models without exposing private data. However, it faces major limitations, including high communication overhead and the need to retrain models when the task is modified.
To overcome these challenges, we propose a privacy-preserving QNN training scheme that utilizes mixed quantum states to encode ensembles of data. This approach allows for the secure sharing of statistical information while safeguarding individual data points. QNNs can be trained directly on these mixed states, eliminating the need to access raw data.
Building on this foundation, we introduce protocols supporting multi-party collaborative QNN training applicable across diverse domains. Our approach enables secure QNN training with only a single round of communication per participant, provides high training speed and offers task generality, i.e., new analyses can be conducted without re-acquiring information from participants.
We present the theoretical foundation of our scheme’s utility and privacy protections, which prevent the recovery of individual data points and resist membership inference attacks as measured by differential privacy. We then validate its effectiveness on three different datasets with a focus on genomic studies with an indication of how it can used in other domains without adaptation.

\end{abstract}

\section{Introduction}
QNNs are gaining increasing interest for their potential to outperform classical neural networks in certain tasks. Leveraging uniquely quantum phenomena such as superposition and entanglement they may be well suited to identify patterns and correlations in data that might be difficult to detect with classical methods \cite{Rist2017,Ciliberto2018,Huang2021,Huang20212,https://doi.org/10.48550/arxiv.2312.03057,Cerezo2022}.

In many fields, such as biomedical research, it is highly beneficial for multiple parties to collaborate by sharing data to jointly train QNN models, as this increases statistical power. However, the sensitive nature of much of this data demands strong privacy protections. Consequently, privacy-preserving quantum computing is essential for secure collaborative analysis.
Blind quantum computing \cite{5438603,Fitzsimons2017} based \ac{QFL} \cite{Li2021,Polacchi2023,Li2024} protocols provide strategies for multiple clients to jointly contribute to QNN model training without exposing private data. However, their practical utility remains limited due to significant computational overhead, particularly from frequent client-server communication required at each optimization step. 

To address these limitations, but also to introduce fundamentally different security schemes to provide robust fallback options in the event that vulnerabilities are discovered in current methods due to unforeseen cryptanalytic advances, we propose a scheme that leverages the non-uniqueness of quantum mixed states \cite{Beltrametti1981} to enable fast, privacy-preserving QNN model training. 
In quantum mechanics, mixed states are probabilistic mixtures of pure states. 
When subjected to completely positive trace-preserving maps, including unitary operations or measurement, the outcomes reflect the weighted average of those obtained from the corresponding pure states. 
Building on this concept, we develop a method for training QNN models using global quantum states—a representative set of mixed states, each constructed from a statistical ensemble of pure quantum states. These global states capture the statistical characteristics of the original input data while concealing individual data points. In our framework, a QNN is trained by minimizing a global loss function, defined as the error between the model’s predictions on these global states and their corresponding global target values. Unlike QFL based approaches, this approach requires only one communication from each participating client and offers significant speed-up by reducing the total number of quantum circuit executions. Based on this framework, we introduce a protocol for privacy-preserving QNN training applicable across diverse domains. A detailed discussion of the additional limitations of \ac{QFL} and a comparison with our approach are provided in \autoref{sec:sec_compare}.

Our scheme satisfies key privacy requirements by eliminating the need for direct access to individual-level data. The non-uniqueness of quantum mixtures \cite{Beltrametti1981} prevents reliable reconstruction of individual compositions from the global quantum state, providing strong protection against composition recovery. Additionally, our approach mitigates the risk of membership inference by effectively "hiding" each individual within a group, offering privacy guarantees similar to those in (quantum) differential privacy \cite{Dwork2013,8049724,10115324}.

Beyond privacy, the utility of global quantum states, e.g., how well they support the intended analysis, is crucial in balancing the privacy-utility trade-off. Within this framework, utility is measured by the QNN’s predictive performance on future individual data points when trained on global states. Below, we highlight three key aspects that offer utility advantages over existing distributed learning schemes. 1) Essential quantum features of the individual states are preserved in the global state, thereby maintaining the utility of the data for effective QNN training. For example, entangling subsystems within a global quantum state corresponds to entangling the same subsystems in each individual state.  2) The representational capacity of mixed states is richer than that of pure states, enabling more expressive modeling. 3) Any mixed state can be represented as a pure state in an enlarged Hilbert space \cite{PhysRevA.73.062309}. Thus, our method effectively embeds multiple data points into a single, higher-dimensional quantum state for downstream model training. This projection mechanism is conceptually aligned with classical techniques such as support vector machines and quantum kernel methods, both of which aim to map data into higher-dimensional spaces to enhance separability and learning performance \cite{10.7551/mitpress/4175.001.0001,Noble2006,PhysRevLett.122.040504,lloyd2020quantumembeddingsmachinelearning}. 
In addition, by reducing the number of quantum states and corresponding QNN evaluations, our approach significantly accelerates the training process, improving the computation efficiency.

We demonstrate the effectiveness of our protocol, both in terms of utility and privacy protection, in the context of genomic research, where collaborative training of neural network models is desirable for increased statistical power, and privacy concerns are especially critical. 
Anonymized genomic datasets, previously assumed to safeguard privacy, have been shown to be susceptible to re-identification, data leakages, and other privacy attacks \cite{Homer2008, Emani2023}, underscoring the necessity of more robust privacy-preserving mechanisms.
Genomic data serves as a unique, immutable identifier for individuals and carries strong predictive power for various health risks. Moreover, because genomic information is shared among family members, privacy breaches can impact not only individuals but also their relatives. Although demonstrated in the context of genomic research, this method is not limited to that domain; it can be directly applied to other fields for privacy-preserving collaborative or delegated QNN training without modification.

This paper is structured as follows:
In \autoref{sec:sec_2}, we present the mathematical foundations of QNN training utility using global quantum states, establish the theoretical basis for privacy protection against composition recovery and membership inference, and introduce our protocol alongside its application to genomic data scenarios. In \autoref{sec:sec_compare}, we provide a more detailed comparison between our method and existing approaches, such as \ac{QFL}.

In \autoref{sec:sec_4}, we evaluate the correctness of our approach and assess its utility by analyzing the performance across different model configurations and benchmarking against both classical and quantum models trained on individual-level data. We also quantify the privacy guarantees of our protocol. \autoref{sec:method} provides the methodological details underlying our approach.

\section{Theoretical Framework}\label{sec:sec_2}
\subsection{Overview}
The overall structure of a gate-based \ac{QNN} typically consists of three main components: 1) an encoder, which transforms classical data into quantum states; 2) a variational circuit, a parameterized unitary quantum circuit used to process the encoded information; and 3) a measurement step, which produces real-valued outputs from quantum states.
In this work, we focus on a specific class of QNNs known as variational quantum classifiers. Variational quantum classifiers \cite{Farhi2018ClassificationWQ,Schuld2020} represent one of the earliest and most well-established applications in quantum machine learning.

In this work, we consider a binary quantum classifier as a test case. The input data are encoded using amplitude encoding $\mathcal{A}$, and RealAmplitudes is used as the ansatz for the variational quantum circuit $\mathcal{U}(\theta)$. The final measurement is performed by evaluating the expectation value of the Pauli-Z operator $P_z$.
The RealAmplitudes ansatz consists of single-qubit rotation gates with two-qubit CNOT gates. This forms a universal set of operations, meaning that any quantum algorithm can, in principle, be implemented using gates from this set. Notably, our scheme is not limited to amplitude encoding; it remains applicable if amplitude encoding is replaced with angle encoding. 

Given a classical input vector $x_u$, where $u$ denotes the index of the data point, amplitude encoding generates a quantum state represented by the density matrix $\rho_u = \mathcal{A}(x_u)$. The variational quantum circuit then transforms this input state into an output state $\sigma_u = \mathcal{U}(\theta) \rho_u\mathcal{U}^{\dagger}(\theta)$. 
The measurement step computes the expectation value of the Pauli-Z operator $P_z$ with respect to the output state, yielding the model prediction:
\begin{eqnarray}
p_u=\text{Tr}(P_{z} \sigma_u).
\end{eqnarray}

\begin{figure}[h]
\centering
\includegraphics[width=1\textwidth]{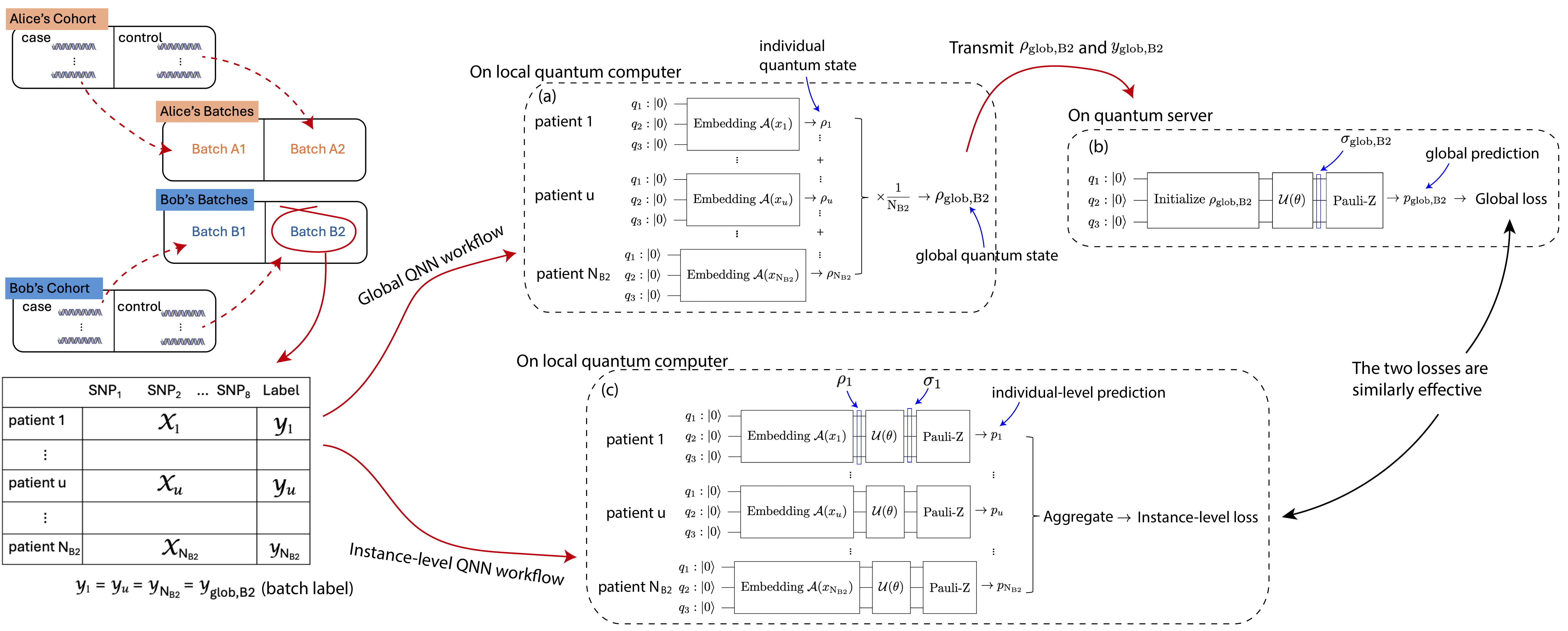}
\caption{Workflow for batching and processing batch B2 with either global QNN or instance-level QNN: (a) construction of the global quantum state, (b) global QNN processing, and (c) instance-level QNN processing.}
\label{fig:illu_all}
\end{figure}

To train the quantum classifier using a classical optimizer, a loss function can be defined based on the expectation values $p_u$.
For example, when using the L1 loss (i.e., the absolute error), the loss function is given by: 
\begin{eqnarray}
\mathcal{L}_{1,\inst}&=&\frac{1}{N_i}\sum_{u=1}^{N_i}|p_u-y_u| \\
&=&\frac{1}{N_i}\sum_{u=1}^{N_i}|\text{Tr}(P_{z} \,\mathcal{U}(\theta)\,\rho_u\,\mathcal{U}^{\dagger}(\theta))-y_u|,
\label{eq:instance_L1}
\end{eqnarray}
where $y_u$ represents the ground truth label. We will refer to this conventional type of loss function as the instance-level loss function and illustrate the workflow of QNNs using this loss function in \autoref{fig:illu_all}\,(c). It is important to note that the instance-level loss function aggregates the loss computed for each individual data point.

Popular optimizers for \ac{QNN}s are typically those suited for optimization problems where the gradient of the objective function is unavailable or costly to compute, due to the lack of efficient gradient measurement algorithms in QNNs.  Many of them, such as COBYLA, rely solely on the average loss across the entire training dataset rather than losses computed for individual data points.

Leveraging this property, along with the fact that both the variational circuit and measurement components (i.e., parts 2 and 3 of the QNN architecture) are linear operations, we propose a method that operates on \textit{global quantum states} representing batches of data points, defined as follows:
\begin{eqnarray}
    \rho_{\g,i}=\frac{1}{N_i}\sum_{u\in C_i}\rho_u 
\end{eqnarray} 
where $C_i$ denotes a batch of data points sharing the same label, and $N_i$ is the number of samples in that batch. For example, $C_0$ consists of data points labeled as signals, while
$C_1$ contains those labeled as controls. The construction of $\rho_{\g,i}$ is illustrated in \autoref{fig:illu_all}\,(a). The global quantum states $\rho_{\g,i}$ serve as representations of the ensemble of original input states and is used during QNN training. 

The global output state, obtained by applying the variational quantum circuit to the global input state $\rho_{\g,i}$ is 
\begin{equation}
\sigma_{\g,i}=\mathcal{U}(\theta)\,\rho_{\g,i} \,\mathcal{U}(\theta)^{\dagger}.
\end{equation}
The measurement output of a global QNN is the expectation value of the Pauli-Z operator with respect to the global output state:
\begin{equation}
    p_{\g,i}=\text{Tr}\Big(P_z\sigma_{\g,i}\Big).
\end{equation}

When training the quantum classifier with global quantum states, the loss function is given by the accuracy of probabilistic predictions, analogous to an L1 variant of the Brier score:
\begin{eqnarray}
\mathcal{L}_{1,\g}&=&\sum_i\big|p_{\g,i}-y_{\g,i}\big| \\
&=&\sum_i\big|\text{Tr}\,(P_{z}\, \mathcal{U}(\theta)\,\rho_{\g,i}\,\mathcal{U}^{\dagger}(\theta))-y_{\g,i}\big|, 
\end{eqnarray}
with $y_{\g,i}$ the label of the batch $C_i$. The index $i$ runs over all batches.
We will refer to this type of loss function as the global loss function. We illustrate the workflow of the QNNs using the global loss function in \autoref{fig:illu_all}\,(b).

One can not trace back the individual data points $x_u \in C_i$ from the global quantum state $\rho_{\g,i}$ because different probabilistic mixtures of individual quantum states can lead to the same global quantum state. However, optimizing the global loss function, which is computed solely from the global quantum state, is as effective as optimizing the instance-level loss function presented in \autoref{eq:instance_L1}.
Henceforth, we will refer to QNN models trained using the global loss function as global QNNs, and those trained with the instance-level loss function as instance-level QNNs.
In \autoref{sec:global_density_matrix}, we provide the mathematical details that demonstrate the effectiveness of the global loss function and thus the utility of the global quantum states. In \autoref{sec:privacy_composition} and \autoref{sec:dp}, we analytically examine the privacy properties of global quantum states with respect to composition recovery and membership inference, respectively.

\subsection{Utility of global quantum states}
\label{sec:global_density_matrix}
Recall the output of the global QNN for batch $C_i$
\begin{equation}
    p_{\g,i}=\text{Tr}\Big(P_z\sigma_{\g,i}\Big).
\end{equation}
And the averaged instance-level QNN output considering all data points in this batch is 
\begin{eqnarray}
    p_{\inst,i}
    =\frac{1}{N_i}\sum_{u\in C_i} \Big(\text{Tr}[P_z\sigma_u ] \Big).
\end{eqnarray}

We notice that  
\begin{eqnarray}
    p_{\g,i}&=&\text{Tr}[P_z \sigma_{\g,i}]=
    \text{Tr}[P_z \mathcal{U}(\theta)\rho_{\g,i} \mathcal{U}(\theta)^{\dagger}]\\
    &=&\text{Tr}[P_z \mathcal{U}(\theta)(\frac{1}{N_{i}}\sum_{u \in C_i }\rho_u) \mathcal{U}(\theta)^{\dagger}]
    =\text{Tr}[P_z \frac{1}{N_{i}}\sum_{u \in C_i }\mathcal{U}(\theta)\rho_u \mathcal{U}(\theta)^{\dagger}]\\
    &=&\frac{1}{N_{i}}\sum_{u \in C_i }\text{Tr}[P_z\mathcal{U}(\theta)\rho_u \mathcal{U}(\theta)^{\dagger}]\\
    &=&p_{\inst,i}
\end{eqnarray}
The above result holds because both the unitary transformation in the QNN and the computation of expectation values are linear operations.

Since the expectation value with respect to the Pauli-Z operator is bounded within the interval $-1<\text{Tr}[\dots] <1$, we rescale it to the range $[0,1]$ using the transformation $(\text{Tr}[\dots]+1)/2$. 
We structure the dataset such that the label $y$ takes binary values in \{0,1\}. Since we have grouped the data points with the same label, the label $y_i$ is identical for all $u\in C_i$.
Either all probability predictions $(\text{Tr}[P_z\sigma_{u}]+1)/2$ for $u\in C_i$ are greater than or equal to $y_i$, or all are less than or equal to $y_i$. Therefore, the rescaled loss for each data point
$$(\text{Tr}[P_z\sigma_{u}]+1)/2-y_i$$ has the same sign across every $u\in C_i$. This uniformity in sign allows us to move the absolute value outside the summation. As a result, the rescaled, instance-level loss averaged over all data points in the batch becomes
\begin{equation}
\mathcal{L}_{1,\text{inst},i}=\frac{1}{N_i}\sum_{u\in{C_i}}|(p_u+1)/2-y_u|=\frac{1}{N_i}|\sum_{u\in{C_i}}[(p_u+1)/2-y_u]|.
\end{equation}

The global L1 loss function for the entire dataset summed over all batches is given by
\begin{equation}
    \mathcal{L}_{1,\g}=\sum_i \Big|(\text{Tr}[P_z\sigma_{\g,i}]+1)/2-y_i\Big|=\sum_i\Big|\frac{1}{N_{i}}\sum_{u \in C_i }(\text{Tr}[P_z \sigma_{u}]+1)/2-y_i\Big|.
\end{equation}
The corresponding instance-level loss for the entire dataset summed over all batches is
\begin{equation}
    \mathcal{L}_{1,\inst}=\sum_i\sum_{i\in C_i}\frac{1}{N_i} \Big|(\text{Tr}[ P_z\sigma_{u}]+1)/2-y_u\Big|.
\end{equation}
The two equations above are equivalent because the L1 loss remains a linear operation when the model output is bounded between 0 and 1, and the true labels are either 0 or 1. Standard neural networks train classifiers over cohorts by optimizing a loss function that aggregates information across all individuals or a suitable approximation. As a result, when using optimizers that rely on the average loss across data points, such as COBYLA, the instance-level loss function conveys the same information during optimization as the global loss function.

We have mathematically demonstrated that, in the case of L1 loss, the global loss function is equivalent to the averaged instance-level loss function. 
Moreover, the global-quantum-state-based approach is also applicable to L2 loss, as well as to cases involving nonlinear postprocessing of the model output, such as applying an activation function instead of the rescaling method discussed earlier. We provide details on the nonlinear cases in the Methods section.

\subsection{Privacy of global quantum states against composition recovery}
\label{sec:privacy_composition}

We aim to quantify the ability to recover the individual constituent states from an observed global quantum state $\rho_{\text{glob}}$. To this end, we estimate the uncertainty that remains about the composition of these underlying individual states after observing $\rho_{\text{glob}}$. This uncertainty is evaluated using conditional entropy as an information-theoretic measure.
For practical purposes, we assume that the individual states considered in our problems come from a finite discrete set. This helps discretize the problem and simplifies the calculation of the quantification of information and uncertainty. Importantly, a higher level of uncertainty about the composition indicates less information leakage, and thus implies a greater degree of privacy.

We start with a finite discrete set of basis quantum states $\{\rho^B_{j}\}_{j=1}^{N_{\text{state}}}$, with $N_{\text{state}}$ the size of the set. This set contains all feasible individual states and the individual states are not necessarily linearly independent.
For example, if each individual state represents a person's SNP vector, and each SNP can take one of three values: $0$, $1$, or $2$, then an individual state is a vector of length $N_{\text{SNP}}$ with entries from $\{0,1,2\}$. The total number of feasible individual states is then $N_{\text{state}} = 3^{N_{\text{SNP}}}$. This set of all feasible SNP conformations can be enumerated as:
\begin{equation}
S^B_1 = (0,0,\ldots,0), \quad S^B_2 = (0,0,\ldots,1), \quad S^B_3 = (0,0,\ldots,2), \quad \ldots, \quad S^B_{N_{\text{state}}} = (2,2,\ldots,2).
\end{equation}
Each such vector $S^B_j$ corresponds to a pure quantum state represented as a density matrix $\rho^B_j$. The density matrix $\rho^B_j$ is obtained via amplitude encoding by first normalizing $S^B_j$, converting it into a state vector $\ket{S^B_j}$ , and then taking the outer product of this state vector with its conjugate transpose $\rho^B_j=\ket{S^B_j}\bra{S^B_j}$.

We also specify a batch size $N_i$, which is the total count of individual data points in the global quantum state.
The global quantum state can be represented as 
\begin{equation}
\rho_\g=\sum_j^{N_{\text{state}}} b_j \rho^B_j,
\end{equation}
where $\mathbf{b}=(b_1,b_2,...,b_{N_{\text{state}}})$ are non-negative integers ($\mathbb{N}^{N_{\text{state}}}$) satisfying $\sum_{j=1}^{N_{\text{state}}} b_{j}=N_i$.
Each coefficient $b_j$ indicates how many times the basis state $\rho^B_j$ appears in the batch. If there exists only a single vector $\mathbf{b}$ compatible with the observed $\rho_{\text{glob}}$, then the individual composition of the batch can be fully inferred, including how many times each SNP conformation occurs. In such cases, privacy is reduced. A large value of $b_j$, i.e., frequent occurrence of a particular SNP conformation, typically reflects population-level information. However, small values of $b_j$, particularly $b_j = 1$, could risk exposing the presence of a specific individual with a particular SNP conformation in the batch, potentially revealing sensitive metadata such as the label associated with that batch.
Conversely, if many different vectors $\mathbf{b}$ are consistent with the observed $\rho_{\text{glob}}$, then it becomes difficult or impossible to infer the exact composition of the batch. In such cases, the appearance of a specific SNP conformation becomes untraceable, thereby ensuring the privacy of individual-level data. 

To this end, we estimate the total number of feasible $\mathbf{b}$ compositions, both with and without the constraint of a fixed global state $\rho_{\text{glob}}$, as a function of the batch size $N_i$ and the number of basis states $N_{\text{state}}$. A larger number of feasible $\mathbf{b}$ vectors, i.e., higher conditional entropy, corresponds to greater uncertainty about the underlying composition, and consequently, a higher level of privacy.

Since the $\rho^B_j$ are not linearly independent, multiple different $\mathbf{b}$ vectors may result in the same $\rho_{\text{glob}}$. We define the set of all feasible such vectors as:
\begin{equation}
B = \left\{ \mathbf{b} \in \mathbb{N}^{N_{\text{state}}} \,\middle|\, \rho_{\text{glob}} = \sum_{j=1}^{N_{\text{state}}} b_j \rho^B_j,\quad \sum_{j=1}^{N_{\text{state}}} b_j = N_i \right\}.
\end{equation}

The size of the set $|B|$ reflects the number of distinct $\mathbf{b}$ consistent with $\rho_{\text{glob}}$, and thus measures the uncertainty in inferring $\mathbf{b}$ from $\rho_{\text{glob}}$.

We first ignore the constraint imposed by $\rho_{\text{glob}}$ and estimate the number of integer vectors $\mathbf{b}$ summing to $N_i$. This can be remodeled into a \textit{Balls into bins problem}, and the total number of feasible combinations is:
\begin{eqnarray}
|B_{all}|&=&\binom{N_i + N_{\text{state}} - 1}{N_{\text{state}} - 1}=\frac{(N_i+N_{\text{state}}-1)!}{(N_{\text{state}}-1)!N_i!}\\
&\sim& \frac{(N_i+N_{\text{state}} - 1)^{N_i+N_{\text{state}} - 1}}{(N_{\text{state}} - 1)^{N_{\text{state}} - 1}N_i^{N_i}}.
\label{eq:lattic_points}
\end{eqnarray}
In the second step, we used the Stirling approximation for the asymptotic behavior of $n!$ for large $N_{\text{state}}$.

We now consider the constraint imposed by fixing $\rho_\g$. Although the components of $\mathbf{b}$ are discrete integers, the construction of density matrices $\rho_j^B$ involves normalizing the corresponding state $S_j^B$.
This normalization allows the components of $\rho_j^B$ to take on a wide range of values, particularly when large $N_{\text{state}}$.
As a result, the global state $\rho_\g$, enriched by the variability in $\mathbf{b}$, can be considered to span a continuous region of the state space rather than being restricted to a discrete, inhomogeneous lattice of points.
Since the $\rho_j$ are linearly dependent, they span only a $d$-dimensional subspace. 
Thus, the constraint $\rho_{\text{glob}} = \sum_j b_j \rho^B_j$ imposes $d$ independent linear conditions, reducing the original $N_{\text{state}}$-dimensional solution space by 
$d$ degrees of freedom. Consequently, the set of feasible $\mathbf{b}$ consistent with both the total count and $\rho_{\text{glob}}$ lies on the intersection of the discrete 
$N_{\text{state}}$-dimensional lattice (as described in \autoref{eq:lattic_points}) with a 
$(N_{\text{state}}-d)$\,-\,dimensional hyperplane.

Hence, we estimate the number of feasible $\mathbf{b}$ vectors to be
\begin{equation}
|B| \approx   |B_{all}|^{\frac{N_{\text{state}}-d}{N_{\text{state}}}}\sim \Bigg[{\frac{(N_i+N_{\text{state}} - 1)^{N_i+N_{\text{state}} - 1}}{(N_{\text{state}} - 1)^{N_{\text{state}} - 1}N_i^{N_i}}}\Bigg]^{\frac{N_{\text{state}}-d}{N_{\text{state}}}}.
\end{equation}
In the example of SNPs, we have $d=N_{\text{SNP}}, \,N_{\text{state}}=3^{d}$.

We can then approximate the remaining uncertainty about the individual composition, i.e., conditional entropy, as:
\begin{equation}
H(\mathbf{b} \mid \rho_{\text{glob}}) \propto \ln |B_\rho| \sim \Big({\frac{N_{\text{state}}-d}{N_{\text{state}}}}\Big) \ln \Bigg[{\frac{(N_i+N_{\text{state}} - 1)^{N_i+N_{\text{state}} - 1}}{(N_{\text{state}} - 1)^{N_{\text{state}} - 1}N_i^{N_i}}}\Bigg].
\end{equation}

This suggests that the number of feasible solutions for $\mathbf{b}$, and hence the uncertainty about the individual composition, increases with larger values of batch size $N_i$ and $N_{\text{state}}$. As a result, inferring the exact composition vector $\mathbf{b}$ from the observed global state $\rho_{\text{glob}}$ becomes increasingly difficult. For instance, even in a relatively small setting with $N_{\text{SNP}} = 4$ and $N_i = 2$, the estimated number of feasible $\mathbf{b}$ configurations exceeds 7000.

\subsection{Privacy of global quantum states against membership inference}
\label{sec:dp}
The membership inference considered in this work is similar to the well-known challenge of determining whether an individual’s data is included in a dataset, either from aggregate statistics or from model outputs under black-box access \cite{Homer2008,10.1145/2976749.2978355,shokri2017membershipinferenceattacksmachine}. 
In our setting, the publicly available information are the global quantum states. Given an observed global quantum state $\rho_{\g}$, we aim to assess whether it is possible to infer if a specific individual state $\rho^{*}$ contributed to the construction of $\rho_{\g}$.
Since batches of quantum data often carry associated metadata (such as class labels with disease status), identifying the presence of a particular individual in the batch may lead to privacy breaches by leaking sensitive information.
Intuitively, as the batch size increases, the influence of any single individual's quantum state becomes increasingly "lost in the crowd" \cite{Bittau2017}—a principle at the heart of many differential privacy mechanisms designed to ensure that no single data point significantly affects the outcome. This is also reflected by the fact the adjacent quantum states, with and without the target individual, become less distinguishable as the batch size grows.

To formalize this intuition, we adopt concepts from quantum differential privacy \cite{8049724}. 
Instead of directly comparing quantum states, we evaluate the classical information obtained through measurements. Specifically, we consider a Positive Operator-Valued Measure (POVM), which extracts classical outcomes from quantum states. In this framework, the global quantum state is a mixture of the target individual state $\rho^{*}$ and other states in the batch. Because the batch composition is unknown to anyone except the trusted local party who owns the data, and can thus be treated as effectively random, the measurement outcomes provide inherent randomness that helps conceal the contribution of the target state $\rho^{*}$.
Here, we establish the formalism and calculate the privacy risk for a fixed $\rho^*$. To quantitatively estimate the privacy risk for a given dataset, one can iterate over all individuals by considering each $\rho^*$ in turn to identify the worst-case privacy risk.

We compare two neighboring global states: one including the individual state $\rho^*$, and one excluding it:
\begin{eqnarray}
\rho_{\g-1}&=&\frac{1}{N_i-1} \sum_{u=1}^{N_i-1}\rho_u,\\
\rho_{\g}&=&\frac{N_i-1}{N_i}\rho_{\g-1}+\frac{1}{N_i}\rho^*.
\end{eqnarray}
We consider here a specific measurement defined by the projection operator corresponding to the individual state $\rho^*$. As we will show later, and as is also intuitively expected, any measurement other than the projection onto $\rho^*$ yields a less distinct signal regarding the presence of $\rho^*$ in the batch.

The probability $p_{\g}^*$ of obtaining $\rho^*$ when performing a POVM defined by the corresponding projection operator of $\rho^*$ on the quantum state $\rho_{\g}$ is
\begin{eqnarray}
    p_{\g}^*&=&\text{Tr}(\rho^*\rho_\g)=\frac{1}{N_i}\text{Tr}(\rho^*\rho^*)+\frac{N_i-1}{N_i}\text{Tr}(\rho^*\rho_{\g-1})\\
    &=&\frac{1}{N_i}\text{Tr}(\rho^*\rho^*)+\frac{1}{N_i}\sum_{u=1}^{N_i-1}\text{Tr}(\rho^*\rho_u)\\
    &=&\frac{1}{N_i}+\frac{1}{N_i}\sum_{u=1}^{N_i-1}\text{Tr}(\rho^*\rho_u),\\
    &=&(\frac{N_i-1}{N_i}p_{\g-1}^*+\frac{1}{N_i})=\alpha p_{\g-1}^*+\beta,
\end{eqnarray}
with $\alpha=\frac{N_i-1}{N_i}$ and $\beta=\frac{1}{N_i}$.
Assuming that all states $\rho_u$ are mutually independent and also independent of $\rho^*$, from the perspective an attacker, the overall overlap $\sum_{u=1}^{N_i-1}\text{Tr}(\rho^*\rho_u)$ can be assumed to be a random variable with the distribution determined by the underlying statistics of the population. 
As a result, $p_{\g}^*$ and $p_{\g-1}^*$ are also random variables with known relationships. If $g_1(p_{\g-1}^*)$ denotes the \ac{pdf} of $p_{\g-1}^*$, 
then the pdf of $p_{\g}^*$ is given by a related pdf $g_2$ via
\begin{equation}
    g_2(p_{\g}^*)=\frac{1}{|\alpha|}g_1\Big((p_{\g}^*-\beta)/\alpha\Big).
\end{equation}

The closer $g_1$ and $g_2$ are, the harder it is to distinguish whether $\rho^*$ was included, thereby implying stronger privacy guarantees. In the quantum $(\epsilon,\delta)$-differential privacy framework \cite{8049724}, the mechanism is considered private if:
\begin{equation}
\Pr[\mathcal{M}^*(\rho_{\g}) \in S] \leq e^{\epsilon^*} \Pr[\mathcal{M}^*(\rho_{\g-1}) \in S]+\delta,
\end{equation}
for all measurable subsets $S$, where $\mathcal{M}^*$ denotes the projection measurement corresponding to $\rho^*$.
The corresponding privacy parameter measuring the level of privacy for a fixed $\rho^*$ is:
\begin{equation}
\epsilon^* =\sup_{p} \left( \ln\Big[\frac{g_2(p)-\delta}{g_1(p)}\Big]\right)= \sup_{p} \left( \ln \left[ \frac{g_1\left( \frac{p - \beta}{\alpha} \right)/\alpha-\delta}{g_1(p)} \right] \right).
\label{eq:epsilon_value}
\end{equation}
For small $\delta$, as $N_i\rightarrow\infty$, we have $\alpha\rightarrow1$ and $\beta \rightarrow 0$, implying $\epsilon^*\rightarrow 0$.  This confirms our expectation: the measurement results on $\rho_\g$ and $\rho_{\g-1}$ become indistinguishable with increasing batch size.
To quantitatively estimate \(\epsilon\) for a dataset without fixing \(\rho^*\) in advance, we iterate over all \(\rho^*\) and compute  
\begin{equation}
\label{eq:final_epsilon_def}
    \epsilon = \max_{\rho^*} \{ \epsilon^* \}.
\end{equation}

Suppose now that the measurement is not perfectly aligned with $\rho^*$, but contains a random term $\rho'=(1-a_{\text{noise}})\rho^*+a_{\text{noise}}\rho_{\text{noise}}$, we have 
\begin{equation}
p_{\g}'=\frac{1-a_{\text{noise}}}{N_i}+\frac{a_{\text{noise}}}{N_i}\text{Tr}(\rho_{\text{noise}}\rho^*)+\frac{1-a_{\text{noise}}}{N_i}\sum_{u=1}^{N_i-1}\text{Tr}(\rho^*\rho_u)+\frac{a_{\text{noise}}}{N_i}\sum_{u=1}^{N_i-1}\text{Tr}(\rho_{\text{noise}}\rho_u)
\end{equation}
and 
\begin{equation}
p_{\g-1}'=\frac{1-a_{\text{noise}}}{N_i-1}\sum_{u=1}^{N_i-1}\text{Tr}(\rho^*\rho_u)+\frac{a_{\text{noise}}}{N_i-1}\sum_{u=1}^{N_i-1}\text{Tr}(\rho_{\text{noise}}\rho_u).
\end{equation}

As before, we can express the relationship as:
\begin{equation}
p'_{\g} = \alpha' p'_{\g-1} + \beta',
\end{equation}
with $\alpha'=\alpha=\frac{N_i-1}{N_i}$ and $\beta'=-\frac{1-a+a\text{Tr}(\rho_{\text{noise}}\rho^*)}{N_i}\leq\beta$.

Since
$\text{Tr}(\rho^*\rho_u)$ and $\text{Tr}(\rho_{\text{noise}}\rho_u)$ are identically distributed, the same base distribution $g_1$ applies. Thus:
\begin{equation}
\epsilon' = \sup_{p'} \left( \ln \left[ \frac{g_1(\frac{p'-\beta'}{\alpha})/\alpha-\delta}{g_1\left(p'\right)} \right] \right).
\end{equation}
Given that $ \beta' < \beta^*$, the argument $\frac{p - \beta'}{\alpha}$ lies closer to $p$ than $ \frac{p - \beta}{\alpha}$. As a result, $g_1\left(\frac{p - \beta'}{\alpha}\right)$ is generally closer to $g_1(p)$ than $g_1\left(\frac{p - \beta}{\alpha}\right)$.
Under the assumptions of small $\delta$ and that $g_1$ changes gradually without significant fluctuations at the scale $\mathcal{O}\left(\frac{1}{N_i}\right)$, as with a Gaussian distribution, we expect $\epsilon' < \epsilon^*$.

In practice, an attack might proceed as follows: the malicious party begins by conducting population studies to prepare. They obtain population statistical information that approximates the distribution of $\rho_u$ and use it to simulate many random $\rho_u$ states. From this distribution, they select random $\rho^*$ states and compute the corresponding distributions for $g_1$ and $g_2$. These precomputed distributions serve as background knowledge. When the attacker receives a global state they wish to analyze, they perform a projection measurement using $\rho^*$, obtain an outcome, and compare it against the two distributions to draw conclusions.

The exact value of the privacy parameter can be estimated using random matrix theory or empirically, provided the distribution of the ensemble $\rho_u$ is known. We will evaluate this value in the results section (\autoref{sec:sec_4}) for the specific dataset used in our experiments.

\subsection{Motivation and Alternatives}
\label{sec:sec_compare}
Alternative to the approach proposed in this work, \ac{QFL} \cite{Chen2021,10988887,10529137,10296523} is one existing approach that enables multiple clients to collaboratively train a global machine learning model without exposing their private data.
The architecture of \ac{QFL} is similar to that of classical federated learning. In a QFL protocol designed for collaborative \ac{QNN} training, each client obtains an initial share QNN as a starting point and uses their own quantum devices to perform local training on their own dataset. Then, instead of sending raw data to a central server, they share only model parameter updates, such as gradient. The central server or a designated client collect the updates and combines them to update the parameters of the shared QNN. The updated QNN model is then sent back to the clients as a new starting point for the next round of training. This process repeats until the model reaches the desired performance. One critical risk for QFL is gradient leakage, where sensitive data can be inferred from shared gradients during training. To protect the privacy of data, additional techniques such as blind quantum computation \cite{Li2021,Fitzsimons2017}, gradient hiding \cite{Li2024} and differential privacy \cite{https://doi.org/10.48550/arxiv.1906.08935} have been introduced to be integrated with QFL schemes. 

A major unresolved limitation of QFL is its high communication overhead, which is further exacerbated by the integration of blind quantum computation. This overhead stems from the frequent client-server communication at each optimization step. Although promising results have been demonstrated on benchmarks such as MNIST and the Breast Cancer Wisconsin dataset \cite{Li2021}, the practical scalability of current QFL implementations remains limited. Existing approaches typically involve fewer than ten qubits \cite{Li2021}, which significantly restricts their applicability to large-scale genomic datasets and other data-intensive domains.
Our approach eliminate the need of frequent client-server communication at each optimization step and requires only one communication from each participating client. 

In addition, current QFL capabilities, along with QNN models more broadly, are limited by the relatively low capacity of Noisy Intermediate-Scale Quantum devices. Our approach addresses this constraint by significantly speed-up by reducing the total number of quantum circuit executions.

Another advantage of our approach is its generality. Unlike many QFL methods that are specifically tailored to a particular use case, our approach is broadly applicable and can be used with any QNN architecture. If a new analysis needs to be performed, the global quantum states can be directly reused without the need to reacquire information from the participating clients who owned the individual level data.

Moreover, \ac{QFL} faces another potential security vulnerabilities include scenarios where an attacker gains control over a participating device and maliciously modifies model parameters which will remain unnoticed because the aggregation procedure prevents the server from inspecting each user’s model update \cite{Kairouz2021}. In contrast, our approach requires trust in only a single party, the central server responsible for training, regarding model security, thereby reducing the attack surface.

Finally, exploring new approaches not only offers better computational efficiency than the existing QFL framework but also introduces fundamentally different security paradigms in the case that vulnerabilities are discovered in QFL due to unforeseen attacks.

\subsection{Protocol}
\label{sec:application_scheme}
\begin{figure}[h]
\centering
\includegraphics[width=1\textwidth]{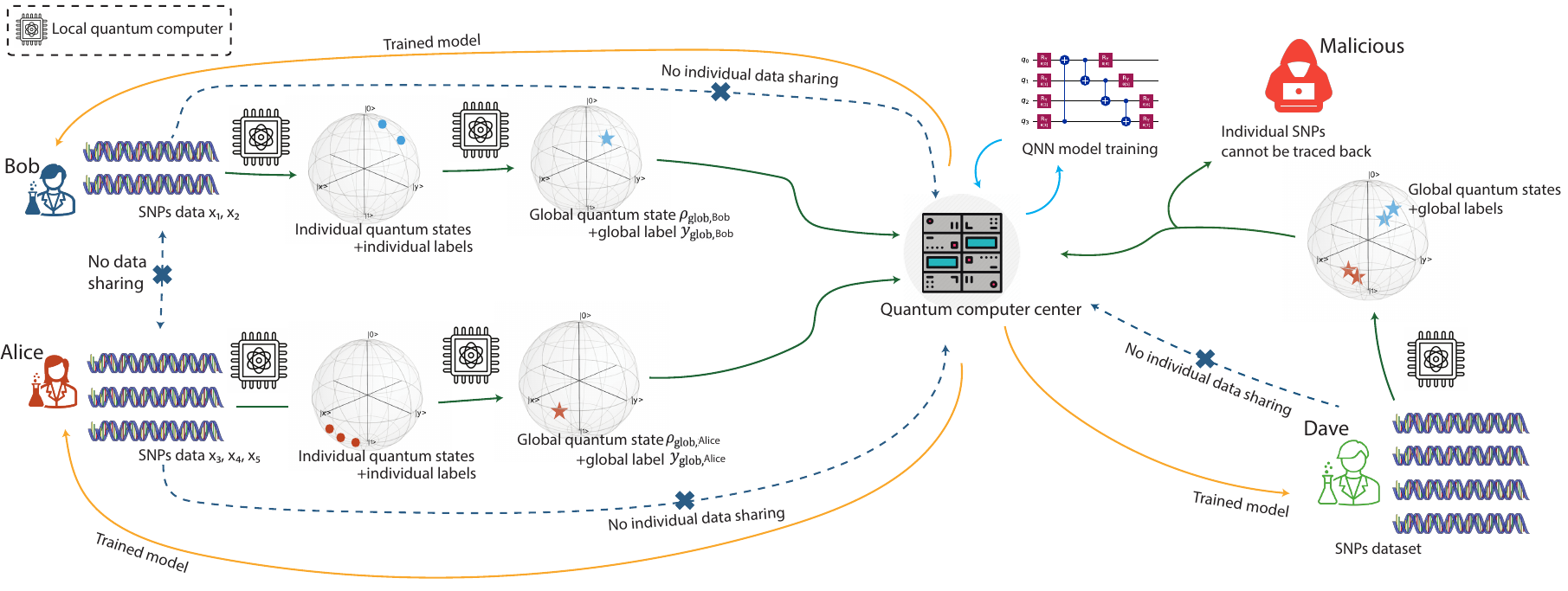}
\caption{Illustration of protocol demonstrating the use of global quantum states for secure private single-party delegated QNN model training and collaborative QNN model training without exposing individual datasets.}
\label{fig:privacy_scheme}
\end{figure}

The effective training of QNN models using global quantum states can be leveraged to develop a quantum-native protocol for private collaborative training, as well as privacy-preserving QNN training on remote quantum servers. We begin by introducing a protocol for single-party delegated QNN training, as it is easier to explain, and then extend it to the multi-party collaborative learning setting. We illustrate practical deployment scenarios in \autoref{fig:privacy_scheme}. For concreteness, we consider classification tasks in which researchers aim to predict whether a given sequence of SNPs is associated with a particular disease. Beyond privacy protection, this protocol also offers potential gains in training efficiency. Since global quantum states summarize batches of data, they reduce the total number of quantum circuit executions required, significantly cutting computation time.

\textit{Single-party delegated QNN training:} We first consider the single-party delegated training scenario as illustrated on the right hand side of \autoref{fig:privacy_scheme}. Dave wants to train a quantum classifier and offload resource-intensive tasks to a remote quantum server while preserving data privacy. 
Using his local, less powerful quantum device, Dave begins by encoding his classical individual SNP data into pure quantum states via a standard quantum encoding technique. He then partitions his training dataset, comprised of data from multiple patients, into batches. Each batch contains multiple patients who share the same disease status but retain distinct SNP profiles. For each batch, Dave constructs a global quantum state using the individual quantum states of the patients in that batch. This results in a "global patient" represented by a mixed quantum state that captures the aggregate statistical features of the group, paired with a corresponding global label indicating the shared disease status.
Next, Dave specifies the desired QNN architecture and transmits both the global quantum states and their associated global labels to the remote quantum server for training. For both the server and any malicious actor who might intercept the global quantum states during transmission, it would be infeasible to reconstruct the original individual data due to the inherent non-uniqueness of quantum mixtures. After the training, the quantum server sends back the trained model to Dave.

\textit{Multi-party collaborative QNN model training:} The multi-party collaborative QNN training scenario is illustrated on the left hand side of \autoref{fig:privacy_scheme}. In this setting, researchers Alice and Bob, working at different institutions, each possess SNPs profiles from patients diagnosed with a specific disease. Their objective is to collaboratively train a QNN model using a remote quantum server capable of handling the resource-intensive training process. But they wish to keep their individual data private from both other researchers and the remote quantum server while ensuring protection against potential security breaches during data transmission. 
Following the same principle as in the single-party setting, Alice and Bob independently preprocess their local data using standard quantum encoding techniques on their own, less powerful quantum devices. Each party then constructs one or more global quantum states that statistically represent subsets of their datasets, with each global state associated with a corresponding global label. 
Once prepared, the global quantum states and their associated labels are sent to the remote quantum server. The server aggregates the inputs from all participating parties and performs QNN training using the specified model architecture. Upon completion, the trained QNN model is returned to all participating entities. At no point does the server gain access to raw individual data, ensuring that privacy is preserved throughout the training process.
Although demonstrated with two clients, the protocol can be easily extended to support multiple contributing parties.

In the experimental sections of this paper, although not explicitly stated, it is assumed that data batching and global quantum state preparation are performed by trusted local parties, while the QNN training on these global quantum states is carried out by a remote quantum server. Throughout the experiments, the entire dataset is treated collectively without explicitly identifying the contributing entities. The multi-party setup can be inferred; for example, in an experiment with 10 batches per class, this could correspond to two participating clients, each independently creating five batches and the corresponding global quantum states, which are then sent to the remote server along with batch size and label metadata.
From the perspective of the remote quantum server, single-party and multi-party training scenarios are similar. In both cases, the server receives quantum states from one or multiple clients without insight into their preparation. It does not even know whether the input states represent individual data points or global quantum states constructed from data batches but simply performs QNN training on the states it receives.

\section{Experimental results}
\label{sec:sec_4}
\subsection{Experimental setup}

To evaluate the correctness of the proposed scheme, we constructed multiple synthetic datasets, including a Pure-State Dataset and a Mixed-State Dataset. Additionally, we generated synthetic SNP datasets and utilized real-world SNP data from the PsychENCODE project to evaluate the accuracy of QNN model training using global quantum states, as well as to assess the utility and privacy of the global quantum states. Detailed descriptions of all datasets are provided in the Methods section. In preparing the SNP datasets in \autoref{sec:result2} and \autoref{sec:result3}, the SNP vectors are ultimately transformed into general floating-point vectors of varying dimensionality. Their resemblance to SNP data does not affect the effectiveness of the global quantum states. As shown in \autoref{sec:result1}, our method is valid regardless of the data’s source or real-world meaning. Consequently, although our privacy and utility evaluations focus on SNP data of classical origin, the results reflect the effectiveness of our approach across diverse domains.

We implemented multiple versions of QNNs. For benchmarking, two QNN models were designed with access to individual data points, each trained using an instance-level loss function. The first, referred to as the \textit{basic QNN}, is based on Qiskit’s original quantum classifier and supports only pure states. The second, the \textit{instance-level QNN}, was custom-developed to handle mixed states and also serves as the foundational implementation for the global QNN model. Finally, we implemented a third model, the \textit{global QNN}, which has access only to global quantum states and is trained using a global loss function. This model simulates the perspective of an remote quantum server. Further details on these models are available in the Methods section.

We employed two batching strategies in this work: random batching and smart batching, as described in the Methods section. 

We evaluate the utility of global quantum states under different settings by first training a QNN on a training set composed of the corresponding global quantum states, and then measuring the QNN’s classification performance, e.g., the test AUC score, on individual data points from the test set. 
To assess privacy in the low statistic regime, we estimate the worst-case privacy parameter $\epsilon$ as  
\begin{equation}
    \epsilon = \max_{p} \left\{ \ln\left( \frac{g_2(p) - \delta}{g_1(p)} \right) \right\},
\end{equation}
which represents a slight modification of \autoref{eq:epsilon_value} and \autoref{eq:final_epsilon_def}. Further details on the privacy risk assessment procedure are provided in the Methods section.

\subsection{Results on synthetic toy datasets}
\label{sec:result1}
We tested all three models (basic, instance-level and global QNN) on the synthetic toy datasets introduced in \autoref{sec:synthetic_quantum_dataset}, and present the results in \autoref{tab:pure_synquantum} and \autoref{tab:mixed_synquantum}. We didn't use any activation but performed a linear re-scaling of the loss function to bring the expectation value of the Pauli-Z operators to the range [0,1] as described in \autoref{sec:global_density_matrix}. When constructing the global quantum states, we used one batch per class, i.e., one global quantum state per class. We found that for both pure- and mixed-state datasets, all three models exhibited nearly identical performance. This suggests that our instance-level QNN implementation is correct and that the global and instance-level objective functions are equivalent if no nonlinear activation function is used.

We found that, although the Pauli-Z expectation value distributions of the two classes are well separated across all values of $e_{\text{shift}}$, making the classification task appear straightforward, the performance of all models remained notably suboptimal. At small $e_{\text{shift}}$, the test AUC is only slightly above 0.5, indicating near-random classification performance. The AUC improves as $e_{\text{shift}}$ increases, or equivalently, as the fidelity between the global quantum states of each class decreases. This suggests that the dissimilarity between the global quantum states is a key factor in classification performance. 
Even when the data points of each class form well-separated clusters, classifiers failed to distinguish between them as long as their global quantum states overlapped. This can occur when data point configurations from different classes result in very similar global quantum states. 

The fact that the well-established implementations of Qiskit quantum classifiers (e.g., Pure-Inst) failed to classify our synthetic toy dataset at small $e_{\text{shift}}$, where the fidelity between the global quantum states of the input classes is high, suggests that careful consideration is needed when assessing whether QNNs are appropriate for a given task.

Our observation implies two important points: 1) before employing quantum classifiers, one should examine whether the global quantum states of the input classes significantly overlap, and 2) the global objective function may effectively replace instance-level objectives. Finally, we note that our synthetic datasets were intentionally designed to test specific aspects of our model. Real-world classification datasets are unlikely to present such particular conditions for quantum classifiers.

\begin{table}
 \caption{Performance of all three models on the pure-state dataset with $e_s=0.4$.}
 \centering
 \scalebox{0.8}{
\begin{tabular}{ c c c c c c c}
\toprule
 $e_{\text{shift}}$ &1.5& 2 &2.5 & 3 &3.5& 4 \\ 
 fidelity &0.835& $0.719$ &$0.608$& $0.528$ & $0.423$ &$0.413$\\ 
 \midrule
 Basic QNN AUC & $0.508/0.503$& $0.629/0.588$&$ 0.697/0.698$& $0.733/ 0.752$ &$0.795/0.798$ & $0.801/0.807$\\
 Instance-level QNN AUC& $0.508/0.503$ &$0.629/0.588$  & $0.696/0.698$ & $0.733/0.751$&  $0.795 /0.798$& $0.801/0.806$ \\
Global QNN AUC&$0.508/0.503$&  $0.629/ 0.588$& $0.696/0.7$& $0.734/0.752$ &$0.795 /0.798$&$0.801/0.806$\\
\bottomrule
\end{tabular}
}
 \label{tab:pure_synquantum}
\end{table}

\begin{table}
 \caption{Performance of all three models on the mixed-state dataset with $e_s=0.2$.}
 \centering
 \scalebox{0.8}{
\begin{tabular}{ c c c c c c c}
\toprule
 $e_{\text{shift}}$ & 0.25 & 0.26 &0.27 &0.28&0.29&0.3 \\ 
 fidelity &  0.816&  0.805& 0.797& 0.789& 0.78& 0.767\\  
 \midrule
 Instance-level QNN AUC& $0.59/ 0.578$&  $0.68/0.655$& $0.784/0.798$& $0.884/0.883$& $0.961/0.967$&  $0.989/0.99$ \\
Global QNN AUC& $0.59/0.578$ & $0.68/0.655$ & $0.784/0.798$& $0.884/0.883$ & $0.961/0.967$ &  $0.989/0.99$ \\
\bottomrule
\end{tabular}
}
 \label{tab:mixed_synquantum}
\end{table}

\subsection{Results on synthetic SNPs dataset}
\label{sec:result2}
We tested instance-level QNN and global QNN with a sigmoid activation function on the synthetic SNPs dataset to demonstrate the effectiveness of using only global quantum states for QNN training. 
We used four layers for the RealAmplitude ansatz. For the instance-level QNN on the 256\,SNPs dataset, we reduced the number of epochs to 30 and patience to 8 due to the long running time.
We varied the number of batches per class from 1 to 50, using both random and smart batching strategies. This corresponds to batch sizes $N_i\in [325,6.5]$.
To explore a wide spectrum of model sizes and complexities, we considered the input feature dimensionality from 16 to 512 SNPs. With amplitude encoding, these input sizes correspond to quantum systems with 4 to 9 qubits.

To evaluate the utility of the global quantum states, we measured the test AUC scores of the QNN models.
For each set of parameters, we conducted five independent runs using a fixed set of five distinct initial weight configurations.
We also tested the logistic regression model as the baseline classical machine learning methods. We used logistic regression for comparison, as it is the core algorithm used in widely adopted Polygenic Risk Score software such as PRSice-2 \cite{Choi2019}. Furthermore, logistic regression shares a similar activation function with the QNN models.

\begin{figure} 
\begin{subfigure}[b]{0.33\textwidth}
\includegraphics[width=1\linewidth]{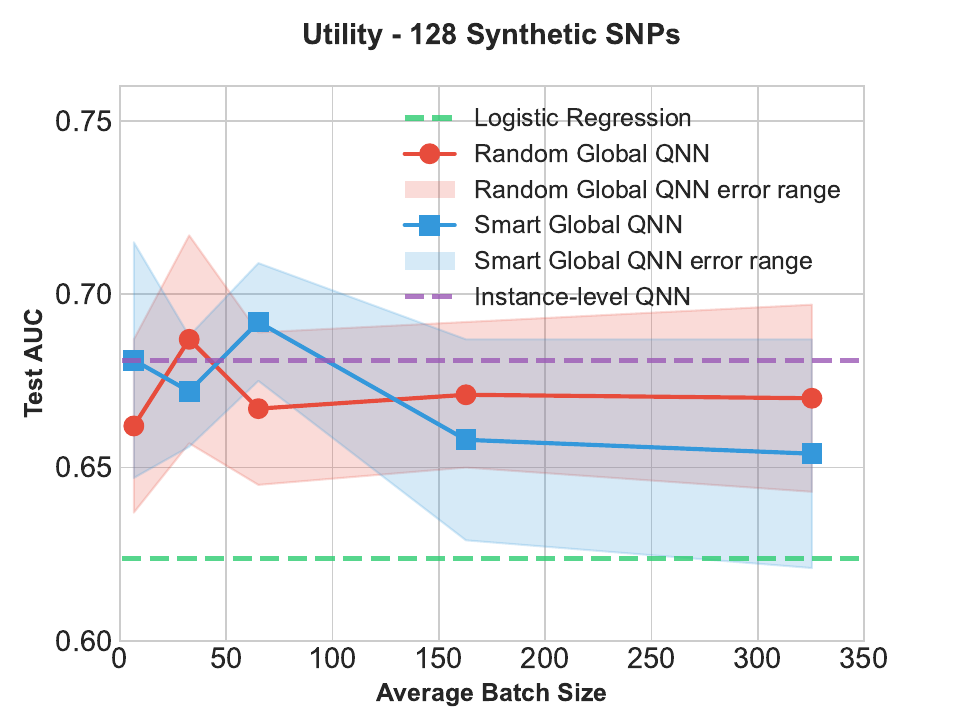}
\caption{}
\end{subfigure}
\begin{subfigure}[b]{0.33\textwidth}
\includegraphics[width=1\linewidth]{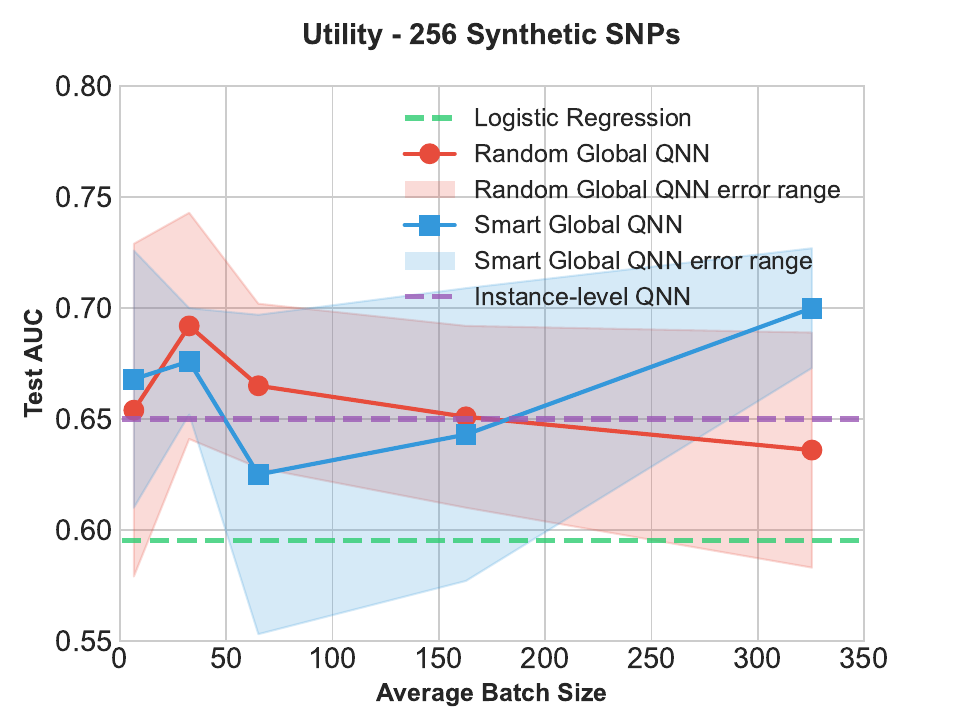}
\caption{}
\end{subfigure}
\begin{subfigure}[b]{0.33\textwidth}
\includegraphics[width=1\linewidth]{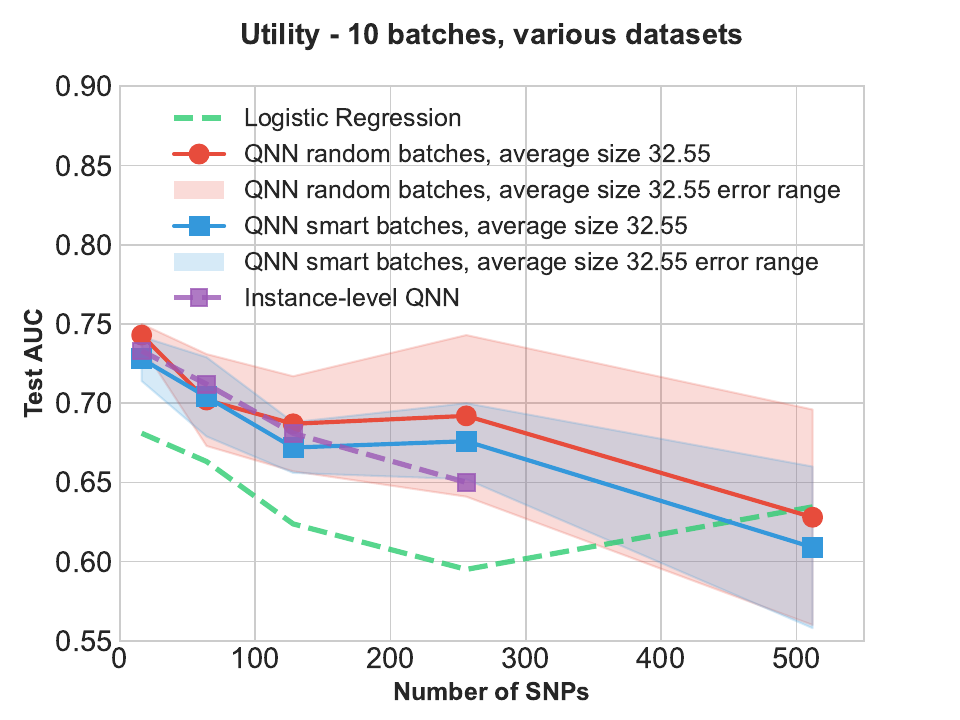}
\caption{}
\end{subfigure}
\begin{subfigure}[b]{0.33\textwidth}
\includegraphics[width=1\linewidth]{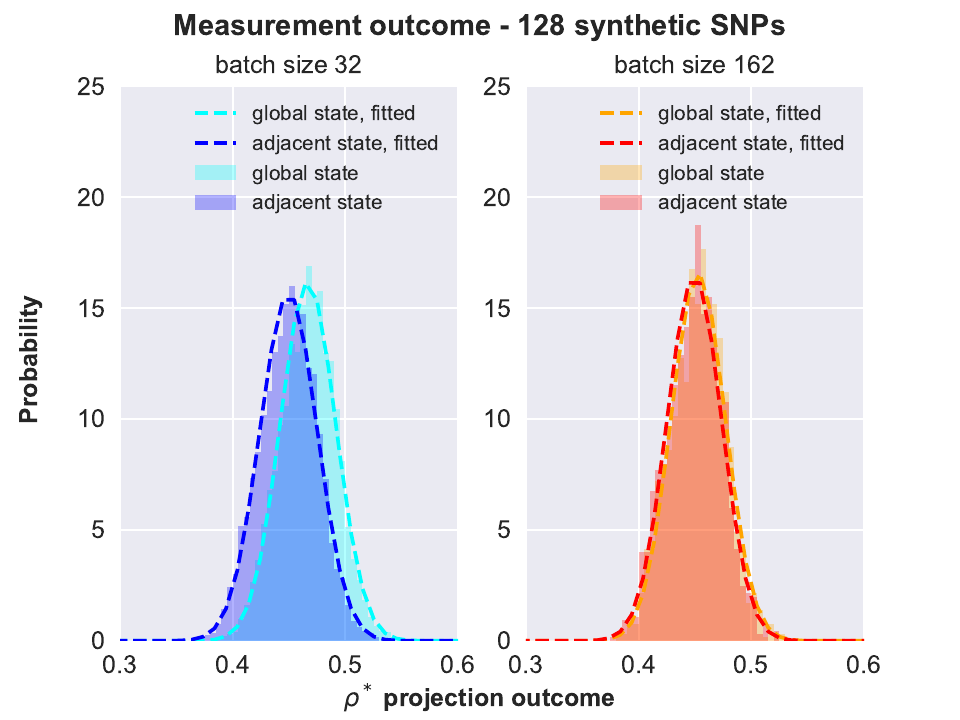}
\caption{}
\end{subfigure}
\begin{subfigure}[b]{0.33\textwidth}
\includegraphics[width=1\linewidth]{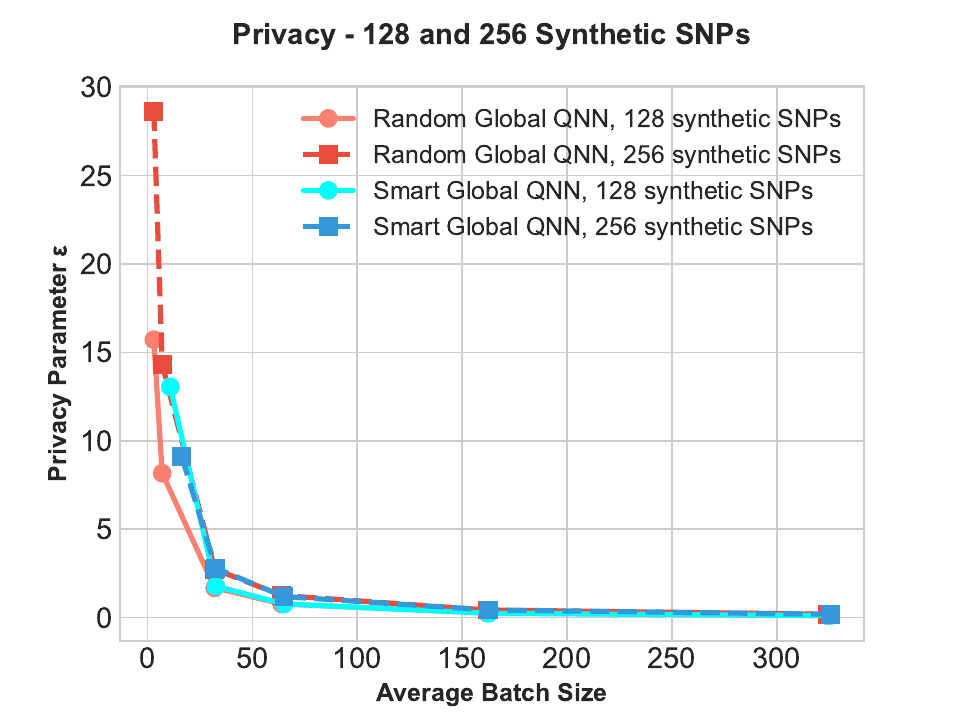}
\caption{}
\end{subfigure}
\begin{subfigure}[b]{0.33\textwidth}
\includegraphics[width=1\linewidth]{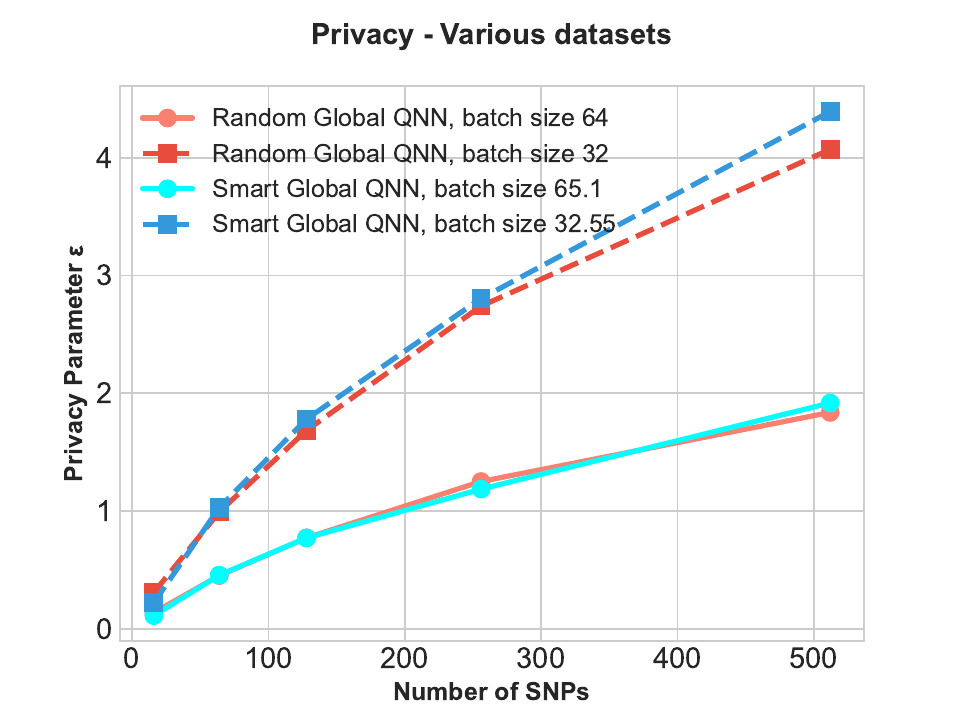}
\caption{}
\end{subfigure}
\caption{Utility and privacy evaluation of global quantum states on synthetic SNPs dataset.
(a), (b) Test AUC performance of instance-level and global QNN models compared to a classical baseline, evaluated on synthetic datasets with 128 and 256 SNPs across varying batch sizes.
(c) Test AUC of global QNN with fixed batch size, instance-level QNN, and classical baseline as input SNP dimensionality varies.
(d) Distribution of $\rho^*$ projection outcomes for random batching.
(e), (f) Privacy parameter $\epsilon$ as a function of batch size and input feature dimensionality.}
\label{fig:synSNPs_AUC}
\end{figure}

In \autoref{fig:synSNPs_AUC}\,(a) and (b), we show the mean test AUC scores for the instance-level QNN, global QNN with varying numbers of batches, and the baseline logistic regression model for two synthetic datasets containing 128 and 256 SNPs, respectively. Shaded areas indicate the error ranges for the global QNN models. 
\autoref{fig:synSNPs_AUC}\,(c) shows the test AUC scores of instance-level QNN, global QNNs with 10 batches, and the logistic regression model for various SNPs dimensionalities. Due to computational limitations, the more resource-intensive instance-level QNN was not evaluated on the dataset with 512 SNPs. 
Together, \autoref{fig:synSNPs_AUC}\,(a)-(c) demonstrate the utility of different global quantum state configurations.
As shown in \autoref{fig:synSNPs_AUC}\,(c), the QNN models consistently outperform logistic regression across datasets with varying SNP dimensionalities, suggesting that QNNs are generally well-suited for classification tasks across the range of input dimensionalities considered.
For the 128- and 256-synthetic SNPs datasets in \autoref{fig:synSNPs_AUC}\,(a) and (b), the test AUC scores of the global QNN fluctuate closely around those of the instance-level QNN across different batch sizes and batching strategies. This indicates a stable utility of the global quantum states as batch size increases and effectively support QNN training.

To quantify the privacy risk, we estimate the privacy parameter $\epsilon$ by evaluating the projection of the target state $\rho^*$ onto the global state $\rho_\g$ (which includes $\rho^*$) and the adjacent state $\rho_{\g-1}$ (which excludes $\rho^*$). For random batching, to improve statistical reliability, we iterate over all individual states and form $\lfloor \text{\#data points} / N_i \rfloor$ non-overlapping batches from the remaining states of the same class. These batches are used to construct $\rho_{\g-1}$, while $\rho_\g$ is formed by adding $\rho^*$ to each batch.
For smart batching, we iterate over all individual states and compute the projection of $\rho^*$ onto $\rho_\g$ (the global state of the batch that contains $\rho^*$) and onto $\rho_{\g-1}$, which is constructed by removing $\rho^*$ from that batch.
This procedure yields a distribution of projection values, as shown in \autoref{fig:synSNPs_AUC}\,(d) for $N_i = 32$ and $162$ under random batching, where $g_1(p^*_{\g-1})$ is labeled as the adjacent state and $g_2(p^*_{\g})$ as the global state.
These distributions are well approximated by normal distributions, and we numerically extract the mean ($\mu$) and standard deviation (std) for both the global and adjacent states.
The overlap (projection) between $\rho^*$ and the other individual states in the batch can be interpreted as the effective noise added for privacy preservation.
The privacy parameter $\epsilon$ is computed as the worst-case privacy loss for a fixed $\delta = 0.05$, using \autoref{eq:epsilon_value}. The parameter $\delta$ serves as a relaxation term, allowing a small probability of the privacy loss exceeding $\epsilon$. This relaxation is particularly important in low-statistics regimes, where rare but large deviations may occur.

\autoref{fig:synSNPs_AUC}\,(e) shows the privacy parameter $\epsilon$ for synthetic datasets with 128 and 256 SNPs, plotted against batch size for both random and smart batching schemes. As observed, $\epsilon$ decreases rapidly as the batch size increases. Around a batch size of 60, the small value of $\epsilon$ less than one indicates a strong level of privacy protection. The random batching scheme generally yields equal or lower $\epsilon$ values compared to smart batching at similar batch sizes. This is because smart batching includes smaller-than-average batches, which yield higher privacy loss due to the worst-case nature of the privacy risk calculation. Moreover, at very small batch sizes, smart batching may form batches containing only a single data point, resulting in infinite privacy loss. Such cases are unsuitable for privacy-preserving computations and are therefore excluded from our analysis. Although smart batching slightly increased privacy risk in this synthetic dataset without showing a clear improvement in utility, this batching strategy is designed to enhance predictive performance while potentially trading off some privacy. Its effects become more apparent in the next section, where we analyze a dataset with a clearer underlying structure. As we will show, smart batching not only demonstrates the utility–privacy tradeoff but can also reduce privacy risk in such settings.

\autoref{fig:synSNPs_AUC}\,(e) presents the privacy parameter $\epsilon$ for synthetic datasets with 128 and 256 SNPs, plotted as a function of batch size for both random and smart batching schemes. As shown, $\epsilon$ decreases rapidly with increasing batch size. At a batch size of around 60, the small value of $\epsilon<1$ suggests a high degree of privacy protection. The random batching scheme provides smaller or similar $\epsilon$ as smart batching for similar batch size, this is due to the presence of smaller batch sizes in the smart batching scheme, since we are calculating the worst-case privacy risk, those smaller than average batches suffer under higher privacy risk. In addition, at a small batch size, smart batching could construct batches containing only one data point, leading to infinite privacy loss. Such scenarios are unsuitable for privacy-preserving computation and are therefore excluded from the results. In \autoref{fig:synSNPs_AUC}\,(f), we plot $\epsilon$ as a function of the number of SNPs (input feature dimensions). The privacy loss $\epsilon$ increases slowly with higher input dimensionality, showing a relatively small effect.

In terms of computational efficiency, using 10 random batches results in each batch containing, on average, 32.5 individual data points, leading to a 32.5-fold reduction in the number of quantum circuit executions compared to instance-level QNN training.

\subsection{Results on PsychEncode SNPs dataset}
\label{sec:result3}

\begin{figure} 
\begin{subfigure}[b]{0.49\textwidth}
\includegraphics[width=1\linewidth]{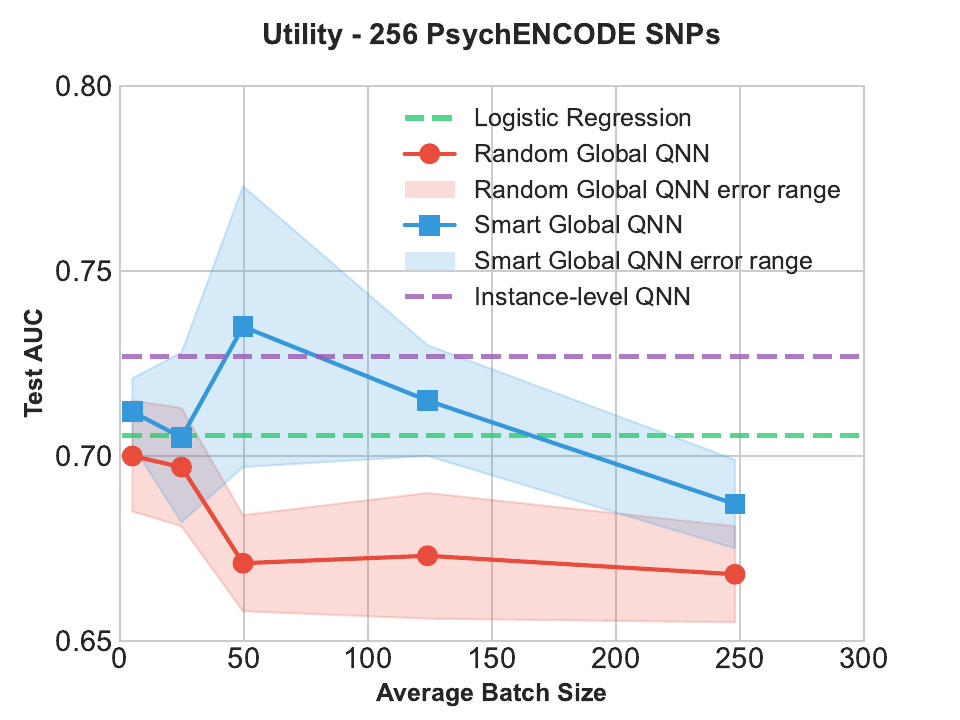}
\caption{}
\end{subfigure}
\begin{subfigure}[b]{0.49\textwidth}
\includegraphics[width=1\linewidth]{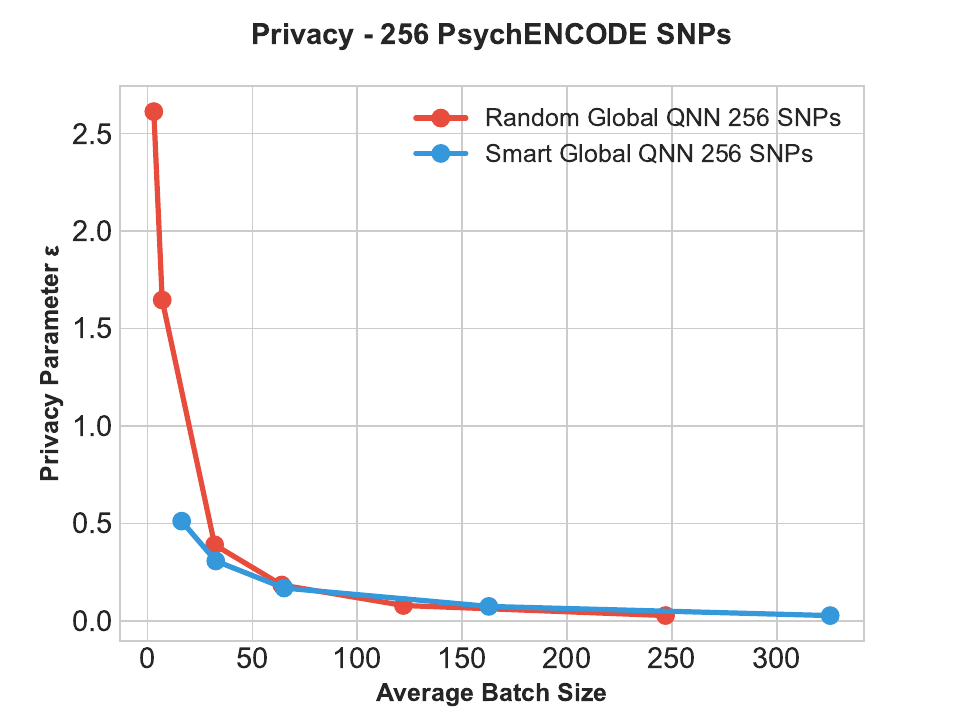}
\caption{}
\end{subfigure}
\caption{ Utility and privacy evaluation of global quantum states on PsychENCODE dataset. (a) Test AUC performance of instance-level QNN, global QNNs across varying batch sizes and the classical baseline models on PsychENCODE datasets containing 256 SNPs. Shaded area is the error range. (b) Privacy parameter $\epsilon$ as a function of batch size.}
\label{fig:PESNPs_AUC}
\end{figure}

We focus on the PsychENCODE dataset with 256 SNPs and associated brain disorder status labels, and employ a QNN circuit with five layers. For baseline comparison, we also evaluate the classical logistic regression model on the same task. 
The test AUC scores of different QNN models and the logistic regression baseline are shown in \autoref{fig:PESNPs_AUC}\,(a), demonstrating the utility of the global quantum states under various settings.
The instance-level QNN outperforms the classical logistic regression, demonstrating the applicability of quantum classifiers for this task. The test AUC of global QNNs with random batching decreases at larger batch sizes and remains sightly below the baseline models. On the other hand, the performance of the global QNNs with smart batching fluctuates around that of the instance-level QNN, demonstrating the utility of the global quantum states constructed via smart batching. 

There is a weak indication,though not statistically conclusive, of a performance peak at an average batch size of approximately 50 for smart batching, followed by a decline for larger batch sizes. 
The performance advantage offered by smart batching over random batching on this real-world dataset could be due to the clearer underlying structure in the real-world data, which facilitates more effective grouping of samples and thus reduces information loss during the non-linear postprocessing. The existence of an optimal batch size suggests that certain grouping strategies better preserve essential data patterns.
In contrast, synthetic SNP datasets, characterized by high randomness and limited inherent structure, did not exhibit performance gains from smart batching.
Overall, QNN models trained on global quantum states offer a robust framework for disease status classification on real-world genomic datasets, demonstrating the practical utility of global quantum states for this task.

\autoref{fig:PESNPs_AUC}\,(b) shows the worst-case privacy parameter $\epsilon$ as a function of the average batch size. The computation of $\epsilon$ follows the same procedure as in the previous section on synthetic data. For batch sizes smaller than 20, smart batching occasionally produces batches containing only a single data point, resulting in infinite privacy loss. Such cases are excluded from the plot. As the batch size increases beyond $\approx 30$, $\epsilon$ rapidly decreases to values below 0.5, indicating strong protection against membership inference.

Smart batching yields slightly lower $\epsilon$ values than random batching, suggesting improved privacy. This improvement may stem from the clearer structure in the dataset, which leads to greater similarity among data points within the same batch. When the target quantum state is hidden among similar states, it becomes harder to distinguish and thus offers better privacy. Despite the presence of small batches due to uneven batch sizes in smart batching, the "hiding among similar states" effect appears to dominate in contrast to the random batching scenario, where such structure is absent.

For this dataset, using 10 batches results in an approximate 25-fold reduction in the number of quantum circuit executions compared to instance-level training.

\section{Discussion}

In this work, we proposed a scheme that uses global quantum states, constructed as probabilistic mixtures of individual data points, to represent the original data ensemble, and demonstrated the effectiveness of using these global quantum states to train QNN models. We established the mathematical foundation showing that optimization on a global loss function defined over these global quantum states simultaneously optimizes the conventional loss function defined over individual data points. We also provided the mathematical basis for the privacy protection offered by our method against both composition recovery and membership inference attacks.
Based on this foundation, we introduced protocols for privacy-preserving QNN training in both single-party delegated QNN learning and multi-party collaborative QNN training settings. Our approach eliminates the need to share individual-level data, ensuring data protection during transmission. Additionally, our method reduces the number of forward-pass executions, thereby accelerating QNN training.
We validated the protocol on quantum binary classification tasks across three datasets: intrinsically quantum data, synthetic genome sequences generated from classical sources, and a real-world genome sequence dataset. We also discussed the general applicability of our method across domains. That is, it could easily be used for other individual level data for instance such as that product marketing or census determination \cite{Abowd_2023,doi:10.1073/pnas.2218605120}.  Overall, our results show that QNN models trained on these global states can match the performance of models trained on individual data points, while offering strong privacy protection and using only a fraction of the computational resources.

\newpage

\section{Methods}
\label{sec:method}
\subsection{Nonlinear postprocessing and nonlinear loss function}
\label{sec:nonlinear}
If a nonlinear operation is involved as postprocessing, for example if an activation function $f$ is applied on the classical output to replace the linear rescaling function, the normalized instance-level loss for data points in a single batch becomes
\begin{equation}
\mathcal{L}_{1,\text{inst},i}=\frac{1}{N_i}\sum_{u\in C_i}|f(p_u)-y_u|=\frac{1}{N_i}|\sum_{u\in C_i} f(p_u)-y_u|.
\end{equation}
The global L1 loss function across the entire dataset is given by
\begin{equation}
    \mathcal{L}_{1,\g}=\sum_i\Big|f(\text{Tr}[P_z\sigma_{\g,i}])-y_i\Big|=\sum_i\Big|f(\text{Tr}[\sum_{u \in C_i }\frac{1}{N_{i}}P_z \sigma_{u}])-y_i\Big|
\end{equation}
while the instance-level loss across the dataset reads
\begin{equation}
    \mathcal{L}_{1,\inst}=\sum_i\sum_{u\in C_i}\frac{1}{N_i} \Big|f(\text{Tr}[ P_z\sigma_{u}])-y_u\Big|.
\end{equation}
To better understand the relationship between these expressions, we examine the loss within a single batch by comparing
\begin{eqnarray}
\Big|f(\text{Tr}[\sum_{u \in C_i }\frac{1}{N_{i}}P_z \sigma_{u}])-y_i\Big| \,(\text{global}) \quad\text{with}\quad
    \Big|\sum_{u \in C_i }\frac{1}{N_{i}}f(\text{Tr}[P_z \sigma_{u}])-y_i\Big| \,(\text{instance-level}).
\end{eqnarray}
Since all data points in a batch $C_i$ share the same label $y_i$, this comparison reduces to analyzing
\begin{eqnarray}
\label{eq:glob_inst_onebatch_activation}
    f(\text{Tr}[\sum_{u \in C_i }\frac{1}{N_{i}}P_z \sigma_{u}])\quad
    \text{and}\quad
    \sum_{u \in C_i }\frac{1}{N_{i}}f(\text{Tr}[P_z \sigma_{u}]).
\end{eqnarray}

These two expressions are not equal when $f$ is nonlinear. 
However, given a monotonically increasing $f$, the two expressions are positively correlated. This suggests that, while the global and instance-level losses differ in magnitude, their gradients will point in the same directions during optimization.

We numerically evaluate the correlation between changes in the global and instance-level losses when transitioning from a quantum circuit with parameters $\theta$ to a new set of parameters $\theta'$, i.e., we assess the correlation between
\begin{eqnarray}
   && \frac{1}{N_i}\sum_{u\in C_i}\Big[ \text{sig}(\text{Tr}[P_z\,\mathcal{U}(\theta')\,\rho_u\,\mathcal{U}^{\dagger}(\theta')])-\text{sig}(\text{Tr}[P_z\,\mathcal{U}(\theta)\,\rho_u\,\mathcal{U}^{\dagger}(\theta)])\Big] \nonumber \\
   && \text{and}\quad
 \text{sig}(\text{Tr}[P_z\,\mathcal{U}(\theta')\,(\sum_{u\in C_i}\rho_u/N_i)\,\mathcal{U}^{\dagger}(\theta')])-\text{sig}(\text{Tr}[P_z\,\mathcal{U}(\theta)\,(\sum_{u\in C_i}\rho_u/N_i)\,\mathcal{U}^{\dagger}(\theta)]).
 \label{eq:sig_global_vs_instance}
\end{eqnarray}
This setup simulates a single step in the QNN optimization process, where a new parameterized circuit $\mathcal{U}(\theta')$ is proposed to replace the current one $\mathcal{U}(\theta)$. During QNN training, a parameter update is accepted if it leads to an improvement in the objective function, i.e., if it decreases the loss in the desired direction.

In our test, each data point $\rho_i$ is a random density matrix. We employ a sigmoid activation function with a temperature parameter $k=10$, i.e., $\text{sig}(x)=1/(1+\exp(-k x))$, consistent with the quantum classifier configuration in the results section. The parameterized quantum circuits $\mathcal{U}(\theta)$ and $\mathcal{U}(\theta')$ are constructed as random circuits by selecting gates uniformly at random from a standard gate set. 

\begin{figure}[H]
    \centering
    \begin{subfigure}[b]{0.49\textwidth}
        \includegraphics[width=1\linewidth]{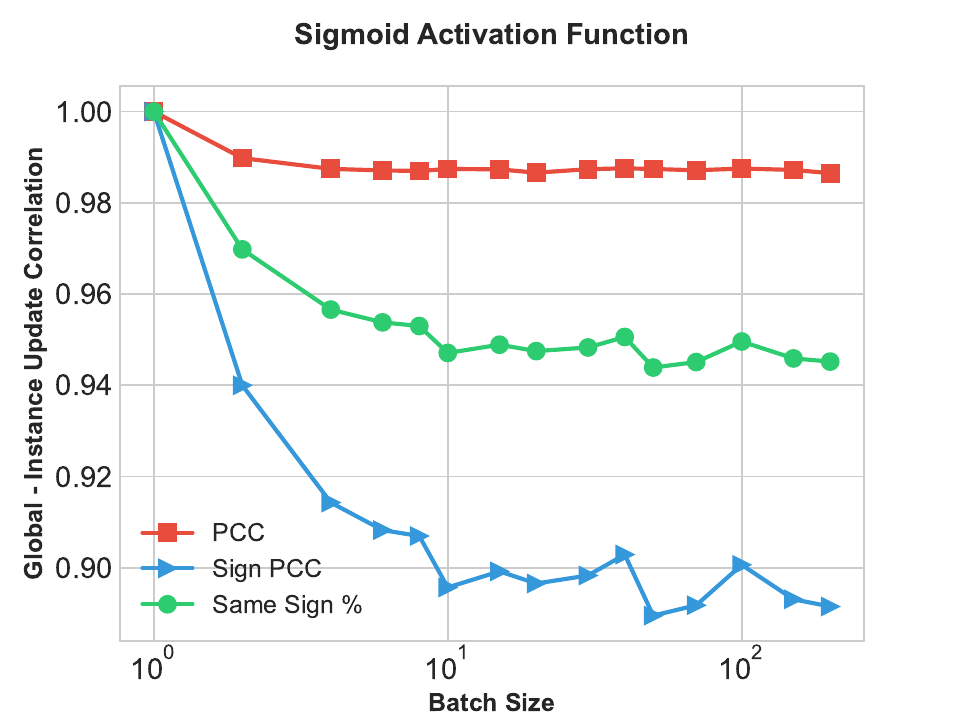}
        \caption{}
    \end{subfigure}
    \begin{subfigure}[b]{0.49\textwidth}
        \includegraphics[width=1\linewidth]{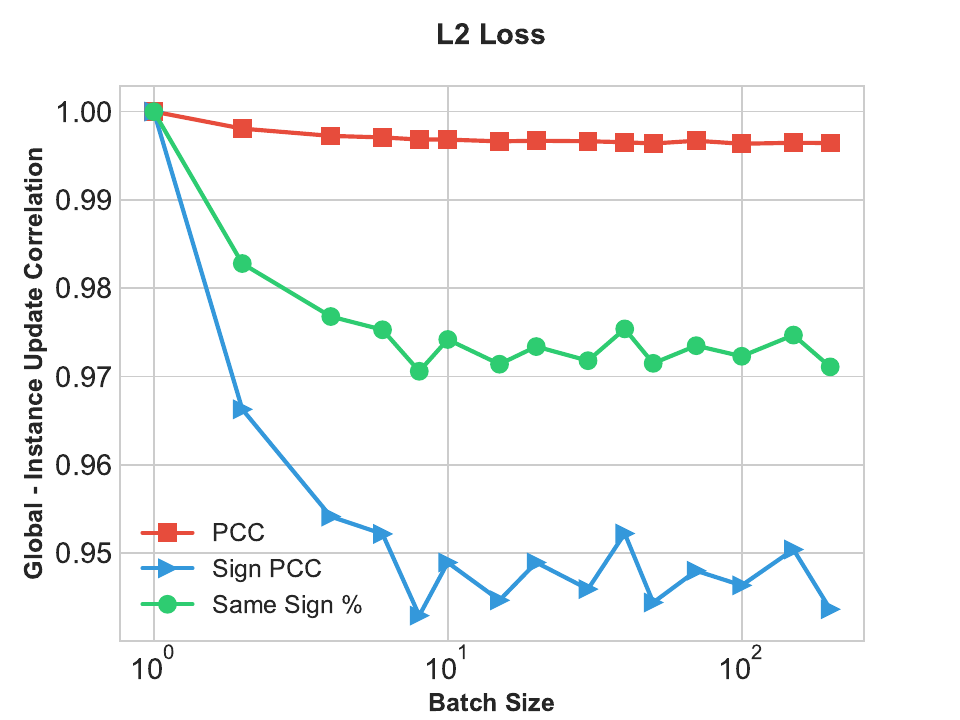}
        \caption{}
    \end{subfigure}
    \caption{Correlation between updates to the global and instance-level objective functions induced by random quantum circuits. We report the \ac{PCC}, the PCC between the signs of the updates, and the percentage of cases where the signs of the global and instance-level updates match for (a) using a sigmoid activation function, and (b) using a L2 loss function.}
    \label{fig:sigmoid_global_instance_test}
\end{figure}

In \autoref{fig:sigmoid_global_instance_test}\,(a), for the scenario of employing a sigmoid activation function and L1 loss, we present the correlation between the global and instance-level expressions from \autoref{eq:sig_global_vs_instance} as a function of batch size $N_{\text{batch}}$, and averaged over 10000 independent trials. 
The \ac{PCC} between the two expressions (red curve) remains consistently high, indicating a strong linear relationship in their responses to circuit updates.
Additionally, the strong correlation between the signs of the changes (blue curve), along with the high probability of agreement of the signs of the change (green curve), indicates that during optimization of the global objective function, an accepted update to the QNN parameters, whether it decreases the loss for class zero or increases it for class one, will induce a change of the same sign in the instance-level objective function. This alignment ensures that the instance-level objective is also optimized in the intended direction.

In the case where an L2 loss function is used instead of L1 loss, the global loss function over the entire dataset reads
\begin{equation}
    \mathcal{L}_{2,\g}=\sum_i\Big(\text{Tr}[P_z\sigma_{\g,i}]-y_i\Big)^2=\sum_i\Big(\text{Tr}[\sum_{u \in C_i }\frac{1}{N_{i}}P_z \sigma_{u}]-y_i\Big)^2
\end{equation}
while the instance-level loss across the dataset reads
\begin{equation}
    \mathcal{L}_{2,\inst}=\sum_i\sum_{u\in C_i}\frac{1}{N_i} \Big(\text{Tr}[ P_z\sigma_{u}]-y_u\Big)^2.
\end{equation}
We again conduct a numerical evaluation of the correlation between changes in the global and instance-level losses induced by a single optimization update, i.e., the correlation between
\begin{eqnarray}
 \label{eq:L2loss_global_vs_instance}
   && \frac{1}{N_i}\sum_{u\in C_i}\Bigg[ \Big(\text{Tr}\Big[P_z\,\mathcal{U}(\theta')\,\rho_u\,\mathcal{U}^{\dagger}(\theta')\Big]-y_u\Big)^2-\Big(\text{Tr}\Big[P_z\,\mathcal{U}(\theta)\,\rho_u\,\mathcal{U}^{\dagger}(\theta)\Big]-y_u\Big)^2\Bigg]  \\
   && \text{and}\quad
 \Bigg(\text{Tr}\Big[P_z\,\mathcal{U}(\theta')\,\big(\sum_{u\in C_i}\rho_u/N_i\big)\,\mathcal{U}^{\dagger}(\theta')\Big]-y_u \Bigg)^2-\Bigg(\text{Tr}\Big[P_z\,\mathcal{U}(\theta)\,\big(\sum_{u\in C_i}\rho_u/N_i\big)\,\mathcal{U}^{\dagger}(\theta)\Big]-y_u\Bigg)^2.\nonumber
\end{eqnarray}
Without loss of generality, we set $y_u=-1$.

We present the \ac{PCC} between the global and instance-level expressions in \autoref{eq:L2loss_global_vs_instance}, the \ac{PCC} between their signs and the probability of both expressions having the same sign in \autoref{fig:sigmoid_global_instance_test}\,(b). The observed high correlation suggests that optimizing the global objective function concurrently improves the instance-level objective function in the same direction when an L2 loss is used.

\subsection{Datasets}
\subsubsection{Synthetic toy dataset}
\label{sec:synthetic_quantum_dataset}
The first dataset we constructed is a synthetic toy dataset aiming to demonstrate the equivalence of the global and instance-level objective functions when no activation function is used, as well as the correctness of our implementation of variational quantum classifier accommodating mixed states. Since this step is solely intended to validate our theoretical framework and algorithm, we use a small system (four-dimensional Hilbert space with two qubits) and directly prepare the density matrices, bypassing the encoding process. It is comprised of two sub-datasets focusing on pure states and mixed states.

\textbf{Pure-state dataset:} We first selected two initial states for each class and then generated the dataset ensemble by perturbing the initial states. For the positive class, the two initial states are 
\begin{equation}
    \ket{x^{\text{ini},1}_a}=\ket{00}\quad \text{and} \quad \ket{x^{\text{ini},1}_b}=\ket{0 1}.
\end{equation}
The initial states for the negative class are 
\begin{equation}
\ket{x^{\text{ini},-1}_a}=\frac{1}{\sqrt{2}}(\ket{00}+\ket{01})\quad \text{and}\quad \ket{x^{\text{ini},-1}_b}=\frac{1}{\sqrt{2}}(\ket{00}-\ket{01}).
\end{equation}
The ensemble is generated by applying a $R_y(\theta_{y1})$-gate on the initial states. $\theta_{y1}$ is a random number drawn from the uniform distribution $\mathcal{U}(-e_s,e_s)$, with $e_s$ a small parameter that controls the overall scale of the perturbation on the first qubit. For data with a positive label, we apply an additional $R_y(\theta_{y2})$-rotation on the second qubit with the random number $\theta_{y2}$ drawn from another uniform $\mathcal{U}(-e_{\text{shift}},e_{\text{shift}})$, with $e_{\text{shift}}$ controlling the difference between the two classes.
For each data point, we stored both its state vector and density matrix expressions for later use, these two expressions are equivalent.

\textbf{Mixed-state dataset:} This dataset simulates intrinsically quantum data, which can only be expressed as density matrices. We use the same initial states as in the pure-state dataset, and generated the ensemble by mixing the initial states with small random density matrices.
The density matrices of the initial states are denoted as $\rho_{\text{ini}}$. We sample random density matrices, $\rho_{\text{random}}$, from the Hilbert-Schmidt metric. For the negative class, the ensemble is generated via
\begin{equation}
    (1-e_{s})\rho_{\text{ini}}+e_{s}\rho_{\text{random}}.
\end{equation}
For the positive class, an additional perturbation is applied, the ensemble is obtained as
\begin{equation}
    (1-e_{s}-e_{\text{shift}})\rho_{\text{ini}}+e_{s}\rho_{\text{random}}+e_{\text{shift}}\begin{pmatrix}
0 & 0 & 0 & 0\\
0 & 0 & 0 & 0 \\
0 & 0 & 0.5 & 0\\
0 & 0 & 0 & 0.5 
\end{pmatrix}.
\end{equation}

Recall the definition of the global quantum state $\frac{1}{N_{i}}\sum_{u \in C_i }\rho_u=\rho_{\g,i}$.
For both datasets, we compute one global quantum state for each class. For instance, $C_0$ could be the negative class and $C_1$ the positive class. The global quantum states of each class are identical when $e_{\text{shift}}=0$ and gradually diverge as $e_{\text{shift}}$ increases. We use fidelity to quantify the distance between the two global density matrices. 
The Pauli-Z expectation values of the data points from the positive class center around 1 and $-1$, while for the negative class, it is centered at zero. 

\begin{figure}[h]
\begin{subfigure}[b]{1\textwidth}
        \centering
        \includegraphics[width=0.325\textwidth]{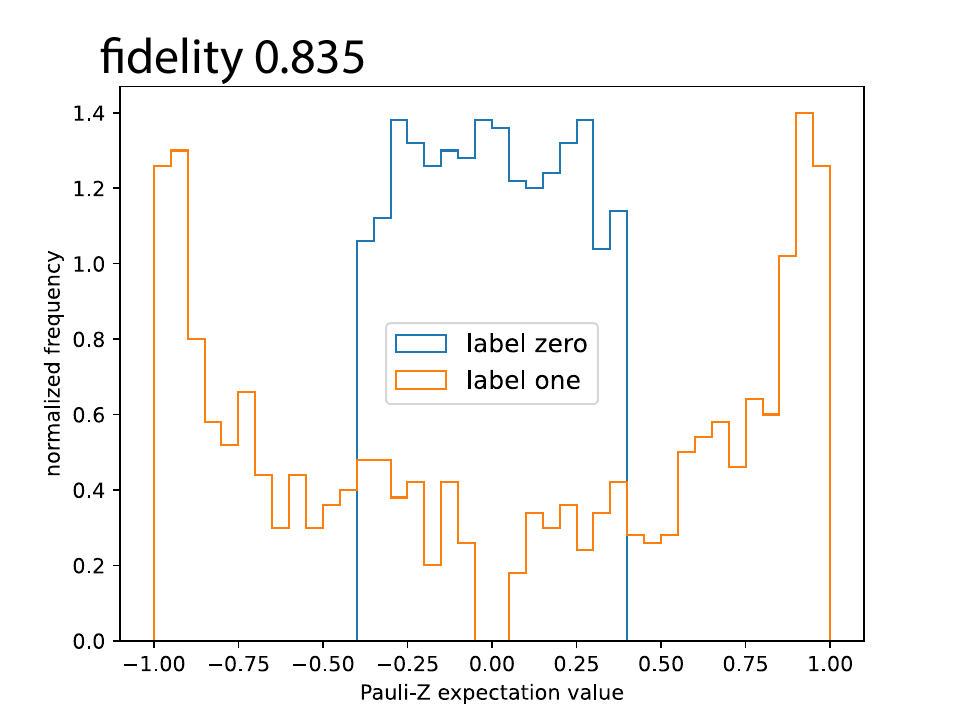}
        \hspace{0.17\textwidth}
              \includegraphics[width=0.325\textwidth]{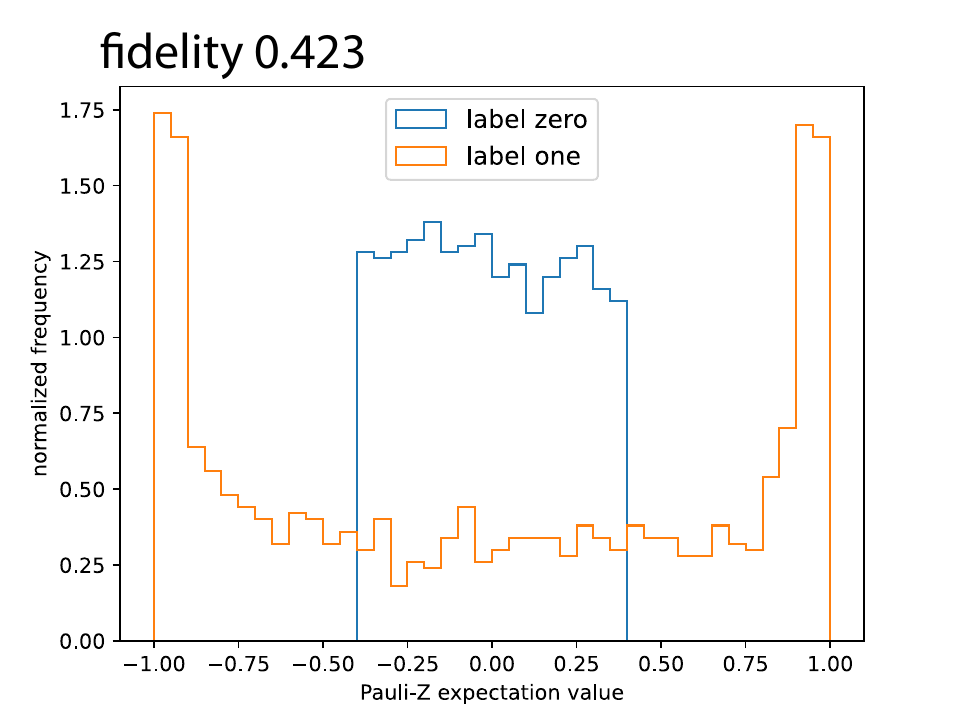}

        \centering
        \includegraphics[width=0.235\textwidth]{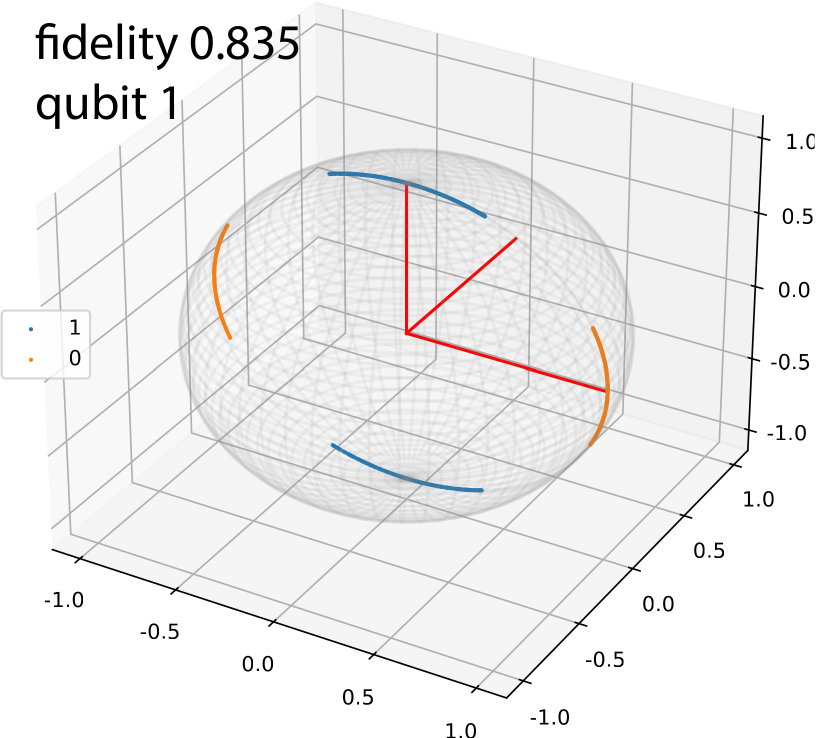}
           \includegraphics[width=0.235\textwidth]{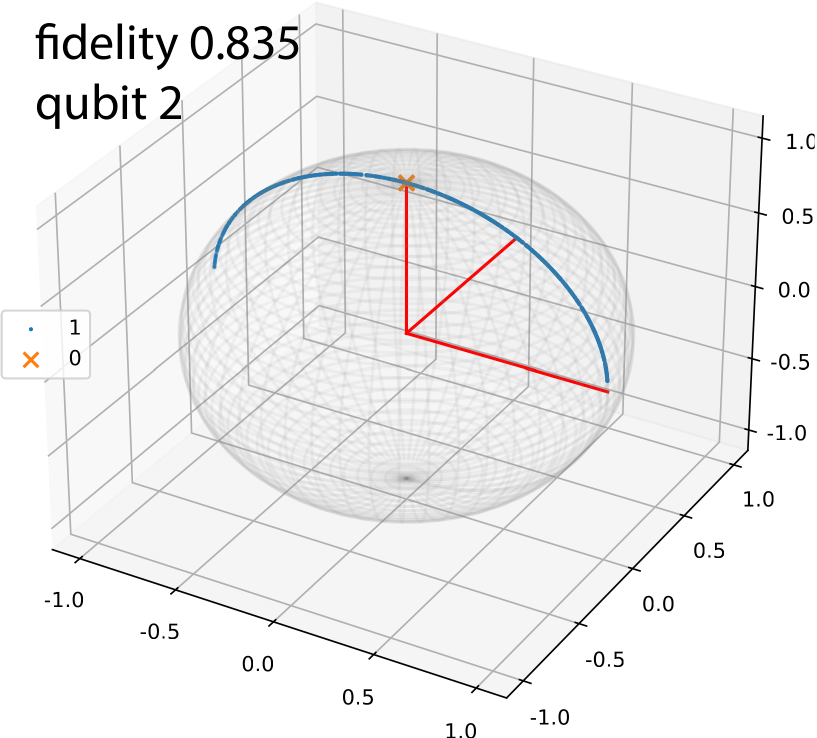}
           \hspace{0.02\textwidth}
              \includegraphics[width=0.235\textwidth]{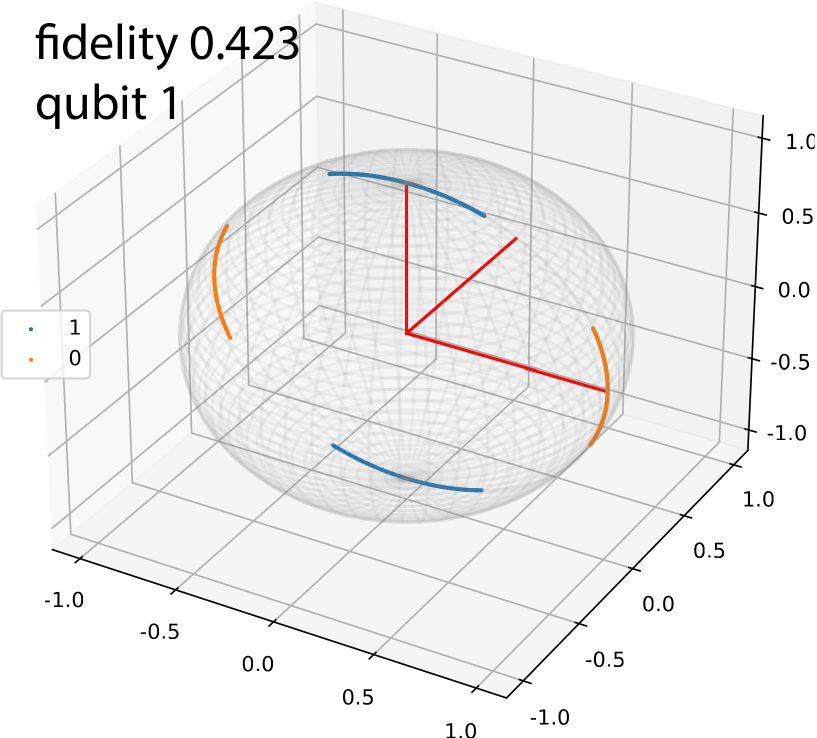}
                \includegraphics[width=0.235\textwidth]{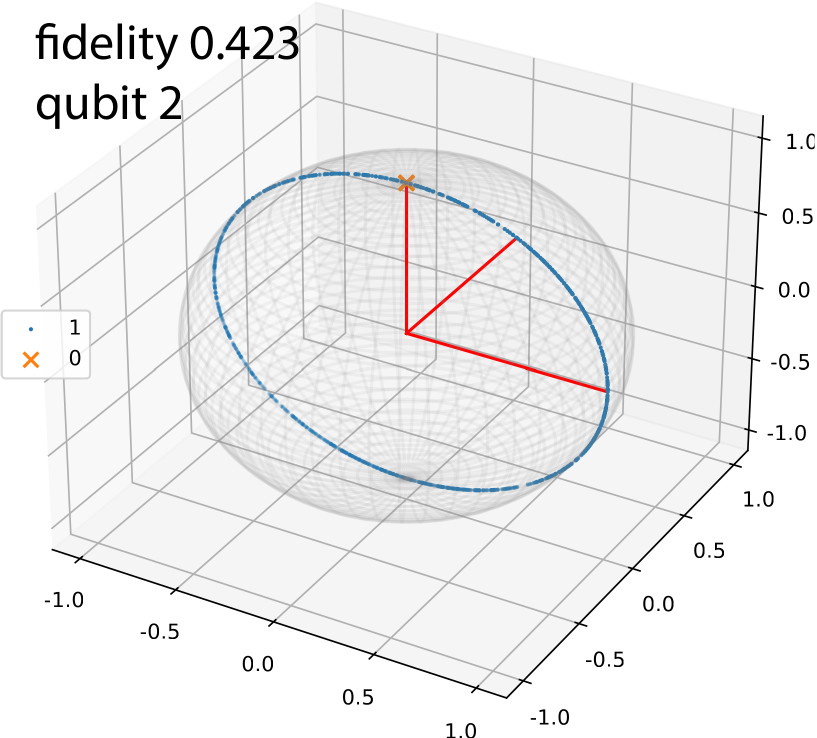}
        \caption{Features of the pure-state datasets with global classes fidelity 0.835 and 0.423. }
    \end{subfigure}
    \begin{subfigure}[b]{1\textwidth}
        \centering
        \includegraphics[width=0.325\textwidth]{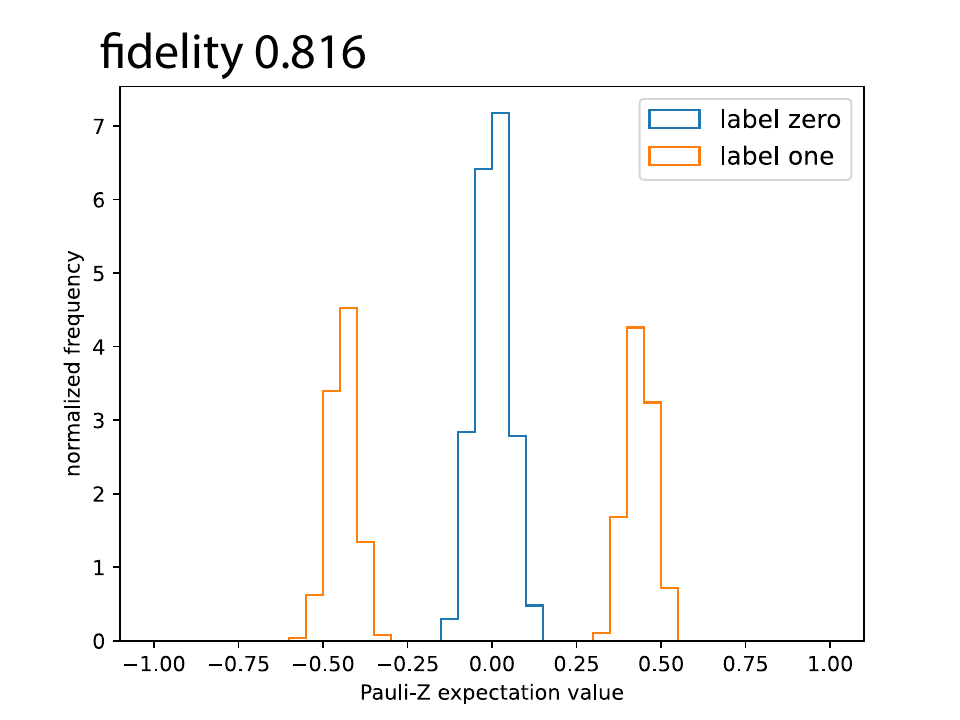}
        \hspace{0.17\textwidth}
              \includegraphics[width=0.325\textwidth]{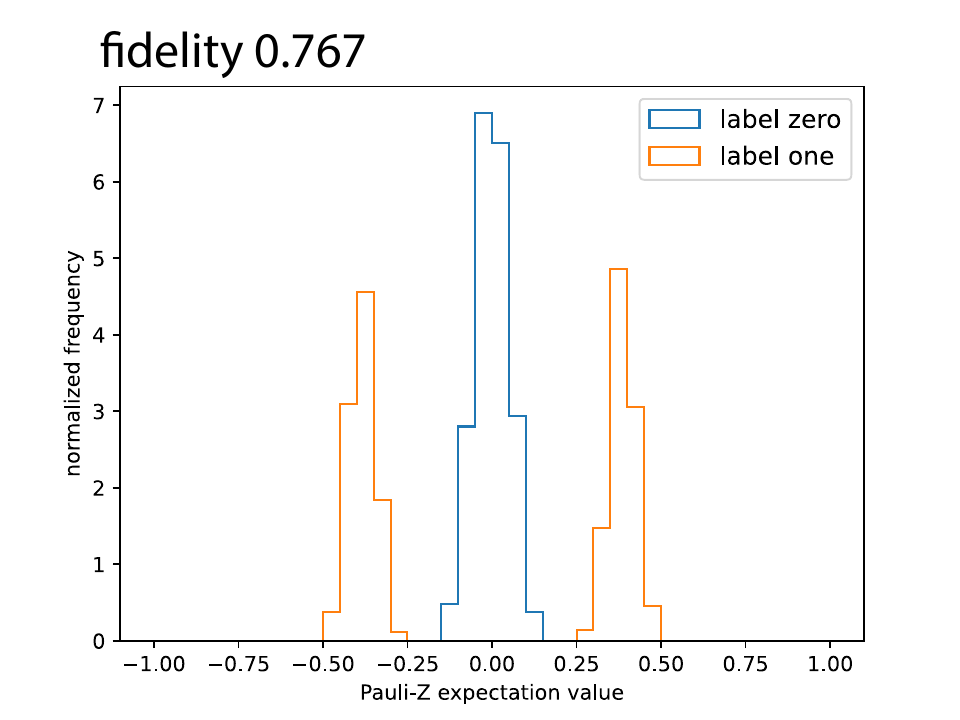}

        \includegraphics[width=0.235\textwidth]{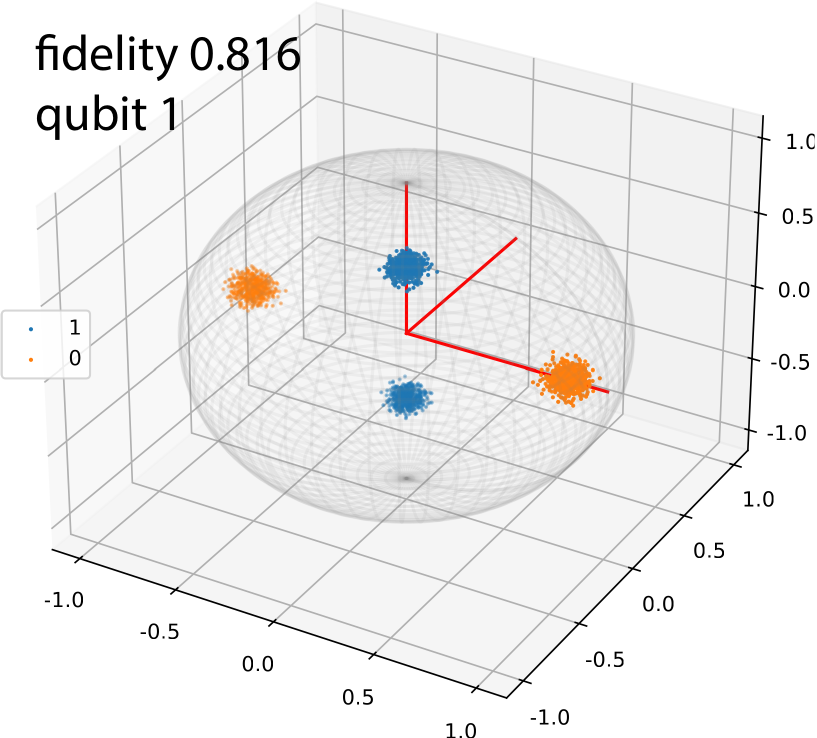}
           \includegraphics[width=0.235\textwidth]{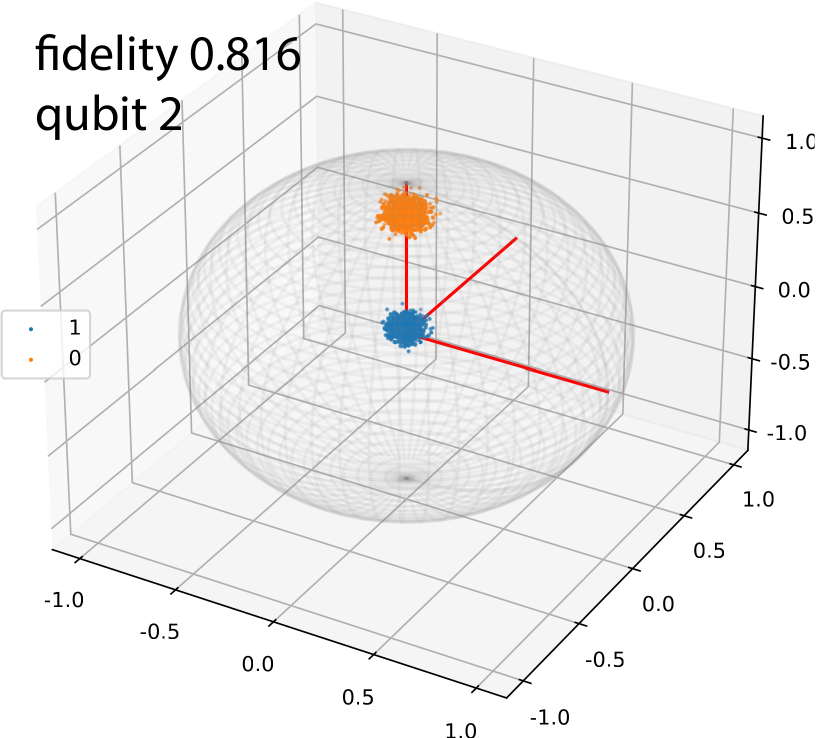}
           \hspace{0.02\textwidth}
              \includegraphics[width=0.235\textwidth]{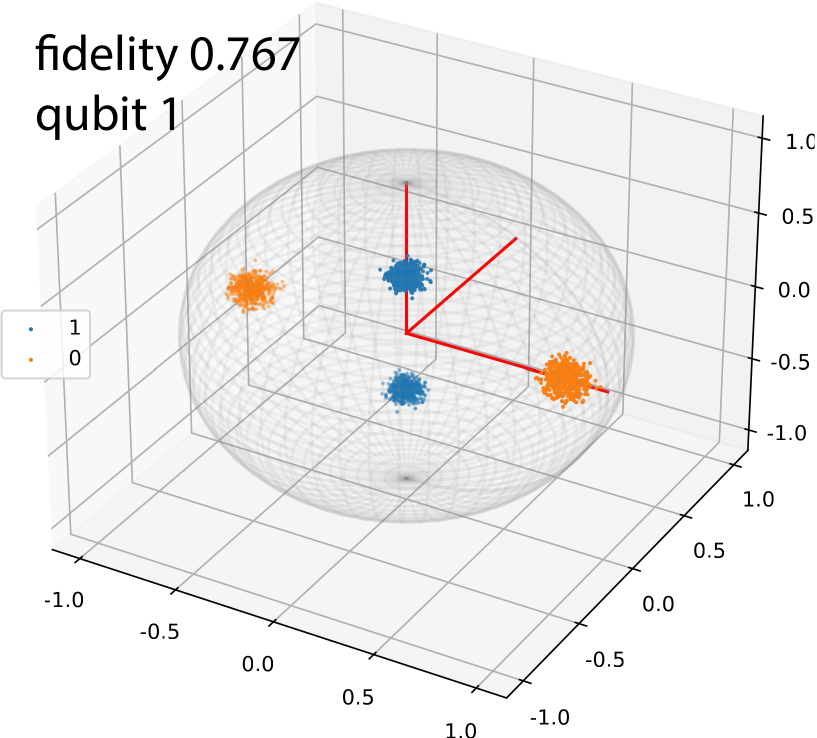}
                \includegraphics[width=0.235\textwidth]{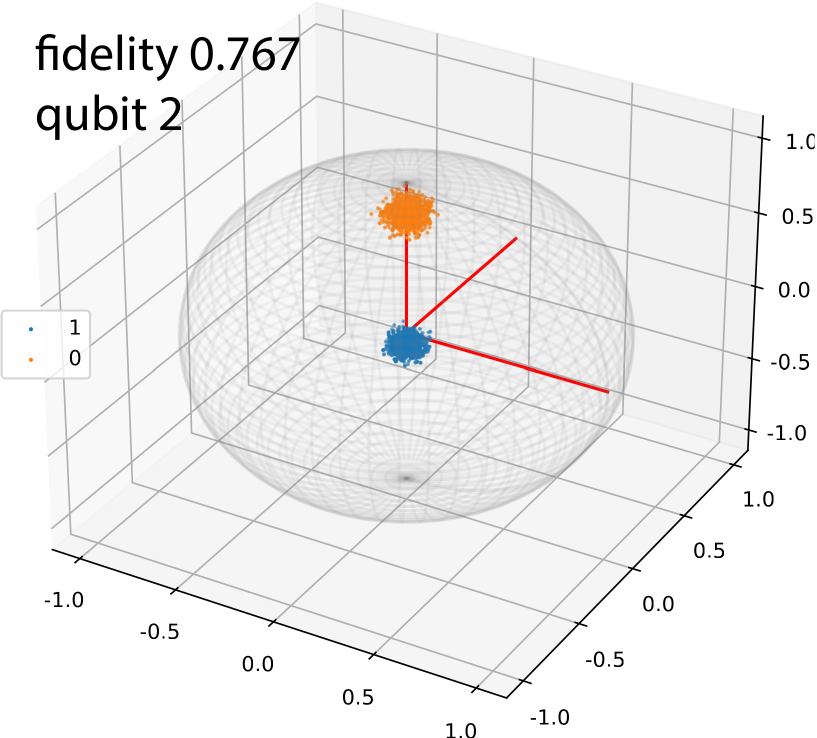}
        \caption{Features of the mixed-state datasets with global classes fidelity 0.816 and 0.767.}

    \end{subfigure}
    \caption{Features of the synthetic toy dataset. Top in each panel: Pauli-Z expectation values. Bottom in each panel: Quantum states distribution in the Bloch sphere.}
    \label{fig:synQuantumDataset_illustration}
\end{figure}

Below in \autoref{fig:synQuantumDataset_illustration}, we present the distribution of the data points (their quantum states) in the Bloch sphere, the distribution of their Pauli-Z expectation values (which is the main classification feature) and the fidelity between the two global quantum states (each representing one class). 

We set the proportion of randomness $e_s$ to a constant and varied $e_{\text{shift}}$ to induce a shift between the global quantum states of class one and class zero.
For the pure-state dataset, we set $e_s=0.4$ and shifted $e_{\text{shift}}$ between 1.5 and 4. For the mixed-state dataset, we set $e_s=0.2$ and shifted $e_{\text{shift}}$ between 0.25 and 0.3. 
Instead of reporting the value of $e_{\text{shift}}$, we reported the fidelity between the global quantum states of the two classes as a measure of their divergence.

\subsubsection{Synthetic SNPs dataset}
To evaluate the effectiveness of QNN training based solely on global quantum states, we demonstrate its application in the context of privacy-preserving genomic data analysis. To this end, we simulate a synthetic case-control dataset at SNP markers using \textit{hapgen2}. \textit{Hapgen2} assumes that disease-associated \ac{SNPs} are in Hardy-Weinberg equilibrium and act independently. To generate phenotypes resulting from interactions between multiple disease-associated \ac{SNPs}, we first use hapgen2 to simulate 500 case and 500 control samples for 10,000 \ac{SNPs} on chromosome 1, to obtain an increased probability of risk alleles in the case samples. We selected six disease-associated SNPs based on the GWAS schizophrenia study with accession number GCST002539; their rsID numbers are rs4648845, rs11210892, rs1702294, rs6670165, rs12129573, and rs140505938.
We artificially increased the heterozygote/homozygote relative disease risks to (1.5/2.25, 1.2/1.44, 1.6/2.56, 2/4, 1.8/3.24, and 1.3/1.69) to enhance the oversampling of haplotypes carrying risk alleles. We disregard the phenotype output of \textit{hapgen2}, as it only simulates independent disease \ac{SNPs}. The increased disease risk was used solely to oversample disease-associated haplotypes. 

To generate interacting phenotypes, we use \textit{SimulatePhenotypes}.
We simulate three phenotype traits based on a two-SNP interaction disease model, referred to as Disease Model 1 in \cite{Marchini2005}. In this model, disease odds increase multiplicatively within and between loci, governed by three parameters: baseline disease odds ($\alpha$), the odds increase for SNP1 ($\theta_1$), and the odds increase for SNP2 ($\theta_2$).
The first two \ac{SNPs} contribute to the first trait with $\alpha = 0.006, \theta_1 = 5, \theta_2 = 5$. The third and fourth \ac{SNPs} contribute to the second trait with $\alpha = 0.004, \theta_1 = 6, \theta_2 = 4$. The last two \ac{SNPs} contribute to the third trait with $\alpha = 0.003, \theta_1 = 7, \theta_2 = 7$. 
These parameters were chosen to generate a high number of cases in the phenotype simulation for genotypes with increased disease-associated haplotypes, while predicting fewer cases in the phenotype simulation for genotypes without increased disease-associated haplotypes.
The final genotypes are encoded as vectors containing 0, 1, and 2, and then normalized. The final phenotype is determined by combining the phenotypes from the three traits: if any of the traits is positive, the data point is considered positive; otherwise, it is classified as negative.
We then balance the dataset using NearMiss\cite{Zhang03}, resulting in 465 disease samples and 465 control samples containing 9993 \ac{SNPs}. To reduce the dimensionality of the input features, we select a subset of \ac{SNPs} with the highest variance across all data points while retaining the disease-associated \ac{SNPs}.

\subsubsection{PsychENCODE SNPs dataset}
We use a dataset of 355 Schizophrenia (SCZ) and 355 Control subjects from the PsychENCODE consortium \cite{emani2024single}, which contains postmortem genomics data from the prefrontal cortex of individuals with psychiatric disorders and matched controls. 
For each individual, we select all single-nucleotide polymorphisms (SNPs) that were associated with 555 key SCZ genes (containing all transcription factors and a set of high-confidence SCZ genes identified in prior studies) as bulk expression quantitative trait loci (eQTLs), as described in \cite{emani2024single}.  This generates a set of 37215 SNPs, which we further reduce to 256 by selecting those showing the strongest Pearson Correlation Coefficient with the binary case/control status in the training set.  Each subject is therefore represented by a compressed genotype vector of length 256, where each entry may be 0, 1 or 2, indicating the number of copies of the alternative allele contained in the subject's genotype for a specific SNP.  

\subsection{Privacy parameter $\epsilon$ computation}
To calculate the privacy parameter $\epsilon$, we iterate over all individual data points, treating each as $\rho^*$. 
For random batches, we randomly select data points from the remainder to construct $\rho_{\g-1}$ and then add $\rho^*$ to form $\rho_{\g}$. For smart batches, the data points are those clustered within the same batch as $\rho^*$.
We projecting $\rho^*$ onto $\rho_{\g-1}$ and $\rho_{\g}$ to obtain $p_{\g-1}^*$ and $p_\g^*$, respectively. 

According to \autoref{eq:epsilon_value} and \autoref{eq:final_epsilon_def}, a single $\epsilon^*$ can be computed for each $\rho^*$, and the overall $\epsilon$ is defined as the maximum of all $\epsilon^*$ values. However, the number of possible batches that can be constructed, and consequently, the number of $p^*_{\g-1}$ and $p^*_{\g}$ values available to compute the distributions $g_1$ and $g_2$ is limited. As a result, especially for larger batch sizes, the estimate of $\epsilon^*$ may be unreliable. To address this, we combine $p^*_{\g-1}$ and $p^*_{\g}$ across all $\rho^*$ to compute aggregate distributions $g_1(p_{\g-1})$ and $g_2(p_{\g})$.
Specifically, we collect all projection outcomes to form distributions of $p_{\g-1}$ and $p_{\g}$. We then fit these distributions with normal probability density functions, obtaining means $\mu_1$, $\mu_2$ and standard deviations $\text{std}_1$, $\text{std}_2$, respectively. The privacy parameter $\epsilon$ is then estimated using:
\begin{equation}
    \epsilon = \max_{p} \left\{ \ln\left( \frac{g_2(p) - \delta}{g_1(p)} \right) \right\},
\end{equation}
where $p$ are 10000 samples drawn from $\mathcal{N}(\mu_2,\text{std}_2)$ and $\delta = 0.05$. Values of $p$ that cause division by zero or the logarithm of a negative number are excluded from the calculation.

\subsection{Batching strategies}
When constructing the global quantum states, the individual input data are divided into batches and one global quantum state is constructed form each batch.
We consider two strategies for dividing input data points into batches. Both strategies are class-specific, e.g., that data points from each class are batched separately. To maintain a balanced dataset, we ensure that each class has the same number of batches thus contributes the same number of global quantum states.
The first approach divides the data randomly into nearly equally sized batches. If the total number of data points is not perfectly divisible by the desired number of batches, leaving a remainder of $N_{\text{remainder}}$, then the first $N_{\text{remainder}}$ batches each receive one additional data point to account for the remainder. The second approach utilized spectral clustering to group together similar data points into a target number of batches. The similarity matrix between data points is computed as the quantum fidelity between them. 

Since density matrices must have unit trace, it is not possible to renormalize each global quantum state to account for the batch size $N_i$ after it is computed as $\frac{1}{N_i}\sum_{u\in C_i}\rho_u$. Thus, the individual data points from a larger batch gets effectively a smaller weight.
To correct this, we add the batch size as weight to the contribution from each global quantum state in the loss function for the smart batching scheme.
For the random batching scheme, it was ensured that the number of data points per batch differs by at most one, which leads to negligible distortion of the original data statistics, especially when the batch size is large. Thus, a batch size correction weight was not used for random batching.

Smart batching is expected to impact the utility and privacy of the global states as follows.

\textbf{Utility:} Without nonlinear loss functions in post-processing (as discussed in \autoref{sec:nonlinear}), both batching strategies yield similar utility. However, when nonlinear operations are involved, smart batching tends to preserve more information and thus results in less utility loss.

\textbf{Privacy:} There are two ways in which smart batching impacts privacy. On one hand, it can reduce privacy protection by producing unbalanced batch sizes, where the smallest batch is smaller than or equal to those in random batching. This size imbalance makes it easier to infer individual contributions within the smaller batches. Additionally, by restricting the space of feasible individual states to those close to $\rho_\g$, smart batching limits the number of possible composition vectors 
$\textbf{b}$, further weakening privacy.
On the other hand, smart batching can improve privacy in datasets with clear structure due to greater similarity among data points within each batch. When the target quantum state is hidden among similar states, it becomes harder to distinguish, thereby enhancing privacy. Ultimately, the net effect of smart batching on privacy depends jointly on the data structure and the batch sizes.

Overall, these two batching strategies provide different trade-offs between utility and privacy, allowing for flexibility depending on the application’s requirements.

\subsection{Code implementation}
\label{subsec:model}
For benchmarking purposes, we implemented two QNN models with access to individual data points (instance-level loss function), alongside a third QNN model with access to only global quantum states (global loss function), reflecting the perspective of a remote quantum server.
\begin{enumerate}
    \item The unmodified Qiskit implementation of EstimatorQNN + NeuralNetworkClassifier. 
    This model can only accommodate pure states and requires all input vectors to be real-valued. Its implementation uses state vectors to represent quantum states and uses an instance-level loss function. We refer to this model as basic QNN.
    \item The EstimatorQNN and NeuralNetworkClassifier classes from Qiskit are modified to accommodate mixed states. We refer to the combination of these two modified base classes as Mixed\_Estimator\_Classifier. Its implementation used density matrices to represent quantum states and employs an instance-level loss function. We refer to this model as instance-level QNN.
    \item The modified EstimatorQNN and NeuralNetworkClassifier classes accommodating mixed states (i.e., Mixed\_Estimator\_Classifier), combined with a global loss function, referred to as global QNN.
\end{enumerate}

We compare basic QNN and instance-level QNN to validate the correctness of our code implementation for Mixed\_Estimator\_Classifier, while the comparison with global QNN tests the effectiveness of our protocol.
As previously noted, optimizers such as COBYLA update parameters based solely on the average loss across all data points. Therefore, when no nonlinear postprocessing, such as activation functions, is applied, all three models are expected to perform nearly identically. When nonlinear elements like a sigmoid activation are introduced, the models’ performance should remain highly correlated. Minor differences may still arise due to numerical discrepancies between state-vector-based and density-matrix-based implementations.

In the computational experiments of this work, we use COBYLA optimizer with maxiter 200.  We set the maximal number of epochs to 50 with a patience of 10 epochs, unless stated otherwise (i.e., training is terminated if the training score does not improve for 10 consecutive epochs). For the RealAmplitude ansatz, we use a \textit{sca} entanglement structure and repeat the basic block, consisting of a rotation layer followed by an entanglement layer, multiple times. This is refered to as the number of layers in the QNN circuit, we treated it as a hyperparameter and selected the best value for it for each dataset.

\printbibliography
\end{document}